\documentclass[sn-basic]{sn-jnl}

\usepackage{subcaption}
\usepackage{lmodern}
\usepackage{float}
\usepackage{adjustbox}
\usepackage{graphicx}

\usepackage{dsfont} 
\usepackage{lastpage}
\usepackage{mathtools}
\usepackage{commath}
\usepackage[nameinlink,capitalise,sort&compress]{cleveref}
\usepackage{caption}

\usepackage{xargs, xparse}
\AtBeginEnvironment{pmatrix}{\everymath{\displaystyle}}

\normalbaroutside

\usepackage{parskip}

\newcommand\bs[1]{\boldsymbol{#1}}

\makeatletter
\newcommand{\algmargin}{\the\ALG@thistlm}   
\makeatother
\algnewcommand{\parState}[1]{\State%
    \parbox[t]{\dimexpr\linewidth-\algmargin}{\strut #1\strut}}

\usepackage[utf8]{inputenc}
\usepackage{pgfplots}
\pgfplotsset{compat=newest}
\usepgfplotslibrary{groupplots}
\usepgfplotslibrary{dateplot}

\newcommand{\expect}{\mathds{E}\expectarg}
\DeclarePairedDelimiterX{\expectarg}[1]{[}{]}{%
  \ifnum\currentgrouptype=16 \else\begingroup\fi
  \activatebar#1
  \ifnum\currentgrouptype=16 \else\endgroup\fi
}

\DeclarePairedDelimiterX{\probinput}[1]{(}{)}{%
  \ifnum\currentgrouptype=16 \else\begingroup\fi
  \activatebar#1
  \ifnum\currentgrouptype=16 \else\endgroup\fi
}

\newcommand{\innermid}{\nonscript\;\delimsize\vert\nonscript\;}
\newcommand{\activatebar}{%
  \begingroup\lccode`\~=`\|
  \lowercase{\endgroup\let~}\innermid 
  \mathcode`|=\string"8000
}

\jyear{2023}%

\begin{document}

\title[Computation of random time-shift distributions]{Computation of random time-shift distributions for stochastic population models}

\author*[1]{\fnm{Dylan} 
\sur{Morris}}\email{dylan.morris@adelaide.edu.au}

\author[1]{\fnm{John} \sur{Maclean}}

\author[1]{\fnm{Andrew J.} \sur{Black}}
\affil[1]{\orgdiv{School of Computer and Mathematical Sciences}, \orgname{The University of Adelaide}, \orgaddress{\city{Adelaide}, \state{SA}, \postcode{5005}, \country{Australia}}}

\abstract{
Even in large systems, the effect of noise arising from when populations are initially small can persist to be measurable on the macroscale. 
A deterministic approximation to a stochastic model will fail to capture this effect, but it can be accurately approximated by including an additional random time-shift to the initial conditions. 
We present a efficient numerical method to compute this time-shift distribution for a large class of stochastic models.
The method relies on differentiation of certain functional equations, which we show can be effectively automated by deriving rules for different types of model rates that arise commonly when mass-action mixing is assumed. 
Explicit computation of the time-shift distribution can be used to build a practical tool for the efficient generation of macroscopic trajectories of stochastic population models, without the need for costly stochastic simulations. 
Full code is provided to implement the calculations and we demonstrate the method on an epidemic model and a model of within-host viral dynamics. 
}


\maketitle

\pagebreak

\section{Introduction}
\label{sec:intro}

The choice of a stochastic or deterministic model for problems in mathematical biology is not always simple. Deterministic models are in general much cheaper to solve, and often the most appropriate if the scale of the system is large, but stochastic effects can still be important on the macroscale \citep{blackStochasticFluctuationsSusceptibleinfectiverecovered2009,butlerFluctuationdrivenTuringPatterns2011,rogersDemographicNoiseCan2012,blackStochasticFormulationEcological2012}.
Examples that motivate this work are populations that initially start from small numbers (cells, virons, invading species, infected individuals, etc.). 
Due to the random nature of the events, these populations initially go through a period of noisy dynamics (and possible extinction), before entering an exponential growth phase \citep{blackEffectClumpedPopulation2014}.
The effect of this early time noise is not averaged out in the large system but instead persists on the macroscale \citep{bakerPersistenceSmallNoise2018}.
For populations that grow to a maximum before declining, as seen in a susceptible-infected-recovered (SIR) model, this noise manifests as variability in the time for the population to peak \citep{nitschkeEffectBottleneckSize2022,curran-sebastianCalculationEpidemicFirst2024}. 
A deterministic approximation to the full stochastic process can accurately capture the large scale dynamics (the early growth rate and shape of the curves at the peak), but fails to capture the variability in the time to peak.

Recent work has shown that the effect of this type of stochasticity on macroscopic population dynamics can be captured by a single univariate random variable representing a time-shift applied to the initial conditions of a deterministic approximation to the full stochastic system \citep{barbourEscapeBoundaryMarkov2015,bakerPersistenceSmallNoise2018,baumanApproximationPopulationsHabitat2023}. 
Although \citet{barbourEscapeBoundaryMarkov2015}  presents the analytical theory for these time-shift distributions and when they are valid, they do not present a method to actually calculate the distributions for general models. In this paper we fill this gap by showing how to numerically compute the time-shift distributions for a broad class of continuous- and discrete-time Markov chain models. Our work therefore greatly expands the applicability of these random time-shift ideas for applied modelling.

The key to calculating the time-shift distribution comes from a branching process (BP) approximation to the full model that is valid at early times. 
The BP approximation can accurately quantify the variability in the population numbers due to the initial stochasticity before the exponential growth phase is entered. 
In the long time limit, a suitably re-scaled version of this process tends to a limiting distribution, which is the distribution of a random variable often denoted $W$ in the literature (\citealp[Chapters III.7 and V]{athreyaBranchingProcesses1972}; \citealp[Chapter~1.8]{modeMultitypeBranchingProcesses1971}; \citealp[Chapter~VI.19]{harrisTheoryBranchingProcesses1964}).
Matching the solution of the BP with a linearised version of a deterministic approximation to the same system shows that the time-shift distribution and $W$ are related via a simple transformation \citep{barbourEscapeBoundaryMarkov2015}. 
Although analytic calculation of $W$ for a few simple models is possible (\citealp{harrisBranchingProcesses1948,harrisMathematicalModelsBranching1951}; \citealp[Chapter~3.1.4]{kimmelBranchingProcessesBiology2015}), in general this appears to be a hard problem, especially for multivariate models. 
Hence we instead develop an efficient numerical scheme to evaluate $W$. 
This is based on a Taylor series expansion of the Laplace-Stieltjes transform (LST) of the distribution, the terms of which can be calculated via differentiation of an implicit functional equation.
Similar to the operation of automatic differentiation routines \citep{bartholomew-biggsAutomaticDifferentiationAlgorithms2000}, we show that this procedure can be encoded as a series of rules for a given model structure. These insights lead to an efficient algorithm and we provide a numerical package to automate this computation for a given model.
We also present another approximate, but even faster, method that employs moment matching, whereby the analytical moments of a surrogate distribution are matched to those of $W$.

Time-shift distributions are useful in themselves for quantifying the role of early stochasticity in population models, but also suggest an elegant and fast method for the generation of macroscopic solutions that also capture this effect. 
This simulation method requires a single solution of the deterministic approximation (typically found by solving a set of coupled ordinary differential equations) that is then replicated many times, and each replicate then shifted in time by a sample from the univariate time-shift distribution. 
This simulation approach can be considered as a type of hybrid simulation method \citep{rebuliHybridMarkovChain2017,kregerHybridStochasticdeterministicApproach2021}, but with a greatly reduced computational cost as only one deterministic trajectory needs to be simulated. 
The reduced computational cost is particularly useful for applications such as Bayesian inference (\citealp[Chapter~IV]{wilkinsonStochasticModellingSystems2019}; \citealp{kregerHybridStochasticdeterministicApproach2021}), where the use of exact \citep{gillespieExactStochasticSimulation1977}, or even approximate \citep{gillespieApproximateAcceleratedStochastic2001}, stochastic simulation methods can become computationally expensive for large systems. 
Moreover, the time-shift distribution can also be used as an importance sampling distribution for even more efficient sequential inference algorithms (\citealp[Chapter~6]{kroeseHandbookMonteCarlo2011}; \citealp{blackImportanceSamplingPartially2018}).

In the next section we present an example to illustrate the basic idea of approximating the macroscopic  stochastic dynamics of a model with a deterministic model subject to random initial conditions. 
We then present our general method for computing $W$, and hence the time-shifts distribution, in Section \ref{sec:methods}.

\section{Example: SIR time-shift distribution}
\label{sec:SIR_example}

We begin with a discussion of the time-shift distribution for an SIR model. The aim of this example is to provide an overview of what a time-shift distribution is and how this arises naturally from an early time analysis of the model.
We choose this particular example as its simplicity allows a  transparent presentation of the main ideas; in addition, analytic expressions for the main results can also be derived, but these are withheld until \Cref{sec:SIR_results}.

The SIR model, in a population of fixed size \(N\), is formulated as a two-state continuous-time Markov chain (CTMC)  where the state of the system at time \(t\) is given by \(\bs{X}(t) = (S(t), I(t))\), and \(S(t)\) and \(I(t)\) are the number of susceptible and infectious individuals respectively \citep{allenPrimerStochasticEpidemic2017}.
An infected individual creates infectious contacts at a rate $\beta$ and if the contact is with a susceptible individual, they become infected. Infected individuals each recover independently at a rate $\gamma$, and hence the mean infectious period is $1/\gamma$.
The possible transitions and corresponding rates are summarised in Table \ref{tbl:SIR_rates}.
We consider a fixed initial number of infectious \(I(0) = I_0\) and hence \(\bs{X}(0) = (N - I_0, I_0)\). 
The regime we are interested in is where the population $N$ is large, but the initial number of infected, $I_0$, is small. 

\begin{table}[!htb]
	\centering
	\begin{tabular}{@{}cccc@{}}
		\toprule
		\(\Delta \bs{X}\) & rate   \\
  \midrule
		\((m,n ) \to (m-1, n+1)\) & $\dfrac{\beta nm}{N - 1}$ \\
		\((m, n) \to (m, n-1)\)   & \(\gamma n\) \\
    \bottomrule
	\end{tabular}
	\caption{Change in state variables and rates for  the CTMC SIR model assuming current state is $X(t)=(m,n)$.}
	\label{tbl:SIR_rates}
\end{table}

For the SIR model a deterministic approximation, valid in the limit $N\rightarrow \infty$, can be derived for the densities $s(t) = N^{-1} S(t)$ and $i(t) = N^{-1} I(t)$ \citep{kurtzLimitTheoremsDiffusion1976}, which are the solutions to the ordinary differential equations (ODEs) \citep{mckendrickStudiesTheoryContinuous1914}
\begin{equation}\label{eq:SIR_det_approx}
    \begin{aligned}
        \dod{s}{t} &= -\beta i s, \\ 
        \dod{i}{t} &= \beta i s - \gamma i.
    \end{aligned}
\end{equation}

\begin{figure*}[!htb]
    \centering
    \includegraphics{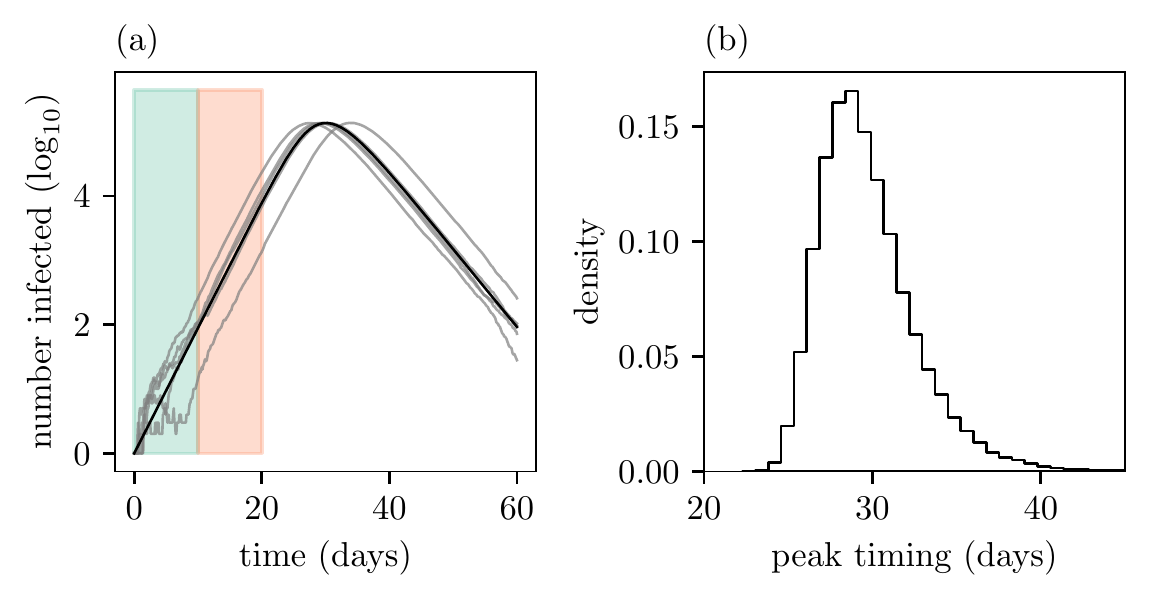}
    \caption{
		Panel (a): Number of infected individuals (\(\log_{10}\)) simulated from the SIR model starting with a single infectious individual in a population of \(10^6\) with model parameters \((\beta, \gamma) = (0.95, 0.5)\).
		Five realisations from the stochastic model, conditional on non-extinction, are shown in grey.
		The solution to the deterministic approximation ($Ni(t)$ where $i(t)$ is the solution of \cref{eq:SIR_det_approx}) is shown in black. 
  The two coloured regions roughly identify the period during which the microscopic stochastic dynamics dominate (green) and the exponential growth phase (red).
		Panel (b): A histogram of the timing of the peak from $10^5$ stochastic simulations.
	}
	\label{fig:SIR_time_delay}
\end{figure*}

Realisations of the stochastic dynamics along with the deterministic solution are shown in \cref{fig:SIR_time_delay}a, where $I_0=1$ and $N=10^6$. 
At early times (green shaded region), the process is strongly affected by stochasticity due to the small numbers of individuals. 
Once the number of infected becomes large enough, the growth becomes exponential (red region), but the random timing of events during the early period affects the transition time at which this occurs.
Over longer time scales, susceptibles become significantly depleted and the non-linearity in the transmission rate becomes important; the exponential growth phase ends and the number of infected peaks and then declines.
What is clear from the realisations is that the stochasicity from the early time dynamics is not averaged out, but persists and is reflected on the macroscale in the random time for the number of infected to peak (\cref{fig:SIR_time_delay}b). For example, a realisation that, by chance, takes a long time to enter the exponential growth phase will also peak much later. 
It can be seen that the deterministic approximation (black curve) captures the macroscopic dynamics well (the exponential growth rate and shape of the curves at the peak), but does not \emph{by itself} capture the stochasticity in the time to peak.

As we shall see in the next two sections, this stochastic behavior at the macroscale can be captured by randomly shifting in time the initial conditions of the deterministic solution. The distribution of this time-shift can be found by equating two early time approximations of the model: a branching process approximation and a linearised deterministic approximation for the mean. Our major contribution in the manuscript is to provide a method to compute this distribution for a general class of discrete- and continuous-time Markov chain models, described in \cref{sec:methods}.

\subsection{Early time approximations to the SIR model}\label{sec:SIR_BP}

The analysis begins by considering the early-time approximation of the SIR model when $N$ is large and $S(0)\approx N$, meaning susceptible depletion can be ignored and the rate of infection is approximately linear, i.e.~$\beta m n / (N - 1) \approx \beta n$. 
Hence, as the rate of recovery is already linear, the number of infected individuals can be approximated with a continuous-time branching process \citep{dormanGardenBranchingProcesses2004}.
Continuous-time branching processes are defined in more detail in \cref{sec:branching_processes}.
The number of individuals infected at time \(t\) is then given by the random variable \(I_b(t)\).  

We next consider two approximations to the branching process dynamics. For the first, define
\begin{equation}\label{eq:W(t)_1D}
    W(t) = e^{-\lambda t} I_b(t) \;,
\end{equation}
a rescaled branching process with initial condition \(W(0) = I_0\). The parameter $\lambda$ is the so-called Malthusian parameter or early growth rate \citep{dormanGardenBranchingProcesses2004}, and for the SIR model is simply $\lambda = \beta - \gamma$ \citep{allenPrimerStochasticEpidemic2017}.
The process \(W(t)\) is well studied in the literature: it is a martingale and converges to a random variable \(W\) almost surely (e.g. \citealp{barbourEscapeBoundaryMarkov2015}; \citealp[Chapter~III.7]{athreyaBranchingProcesses1972}). Let us pause to interpret the limiting behaviour when the time is long enough such that $W(t) \approx W$. In this limit, the number of infected $I_b(t) \approx e^{\lambda t} W$ only depends on $t$ in the exponent, and hence the dynamics have entered the exponential growth phase. 
Substituting this approximation in \cref{eq:W(t)_1D} and rearranging, we have, 
\begin{equation}\label{eq:SIR_stochastic_sol}
  I_b(t) \approx \exp{\left(\lambda \left(t + \lambda^{-1}\log{W}\right)\right)}.
\end{equation}
In this equation, $W$ has been placed in the exponent so that we can in future interpret it as a shift to the time $t$.

A second approximation to the BP dynamics is to just consider the mean number of infected, \(I_d(t) = \expect{I_b(t)} = I_0 e^{\lambda t}\) \citep[Chapter~III]{athreyaBranchingProcesses1972}. 
This can be expressed similarly to \cref{eq:SIR_stochastic_sol}
\begin{equation}\label{eq:SIR_mean_sol}
  I_d(t) = \exp{\left( \lambda \left(t + \lambda^{-1}\log{\expect{W}}\right) \right)} 
\end{equation}
where \(\expect{W} = I_0\) since \(W(t)\) is a martingale (\citealp[Chapter~III.7]{athreyaBranchingProcesses1972}; \citealp{barbourEscapeBoundaryMarkov2015}).
The two approximations for the dynamics look similar, and clearly capture the observed exponential growth in this model, but are different. The approximation for $I_b(t)$ is stochastic as $W$ is a random variable, $I_d(t)$ is deterministic.

\subsection{Random time-shifts}\label{sec:random_time-shifts_SIR}

\begin{figure*}[!htb]
    \centering
    \includegraphics{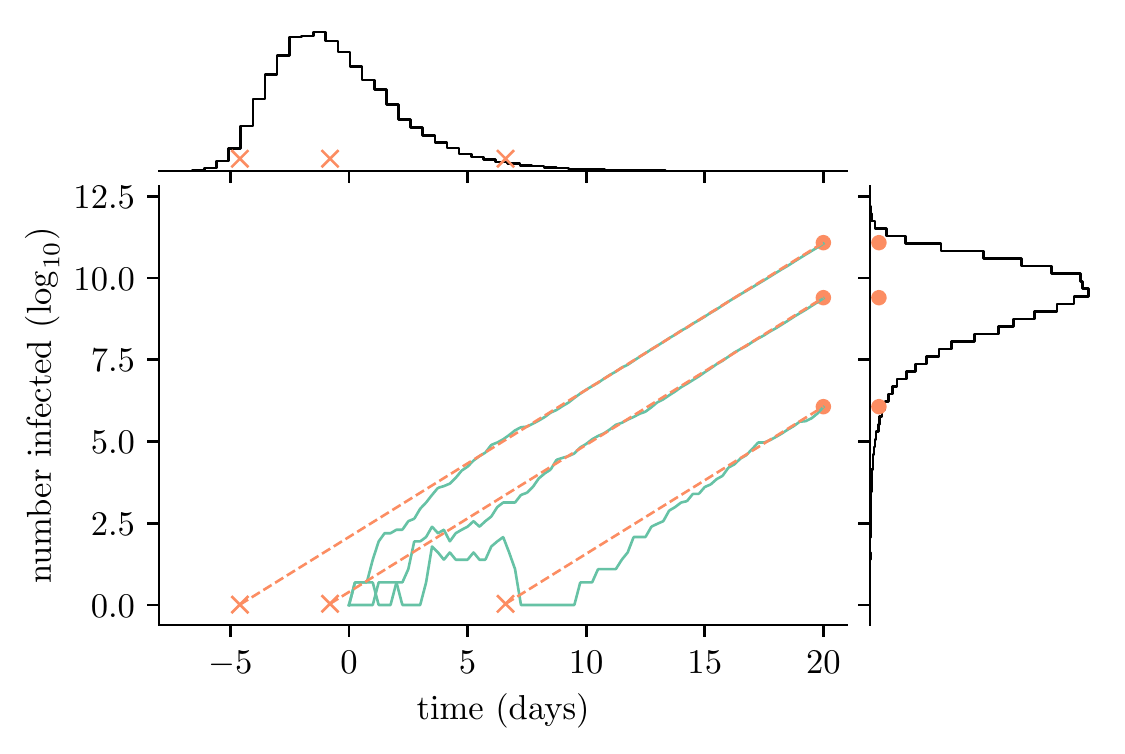}
    \caption{
        Three realisations of the stochastic SIR model are shown in green, plotted on a log scale, with $I_0=1$.
        The orange dashed lines are the projections down from the point at which the stochastic trajectories approximately follow the deterministic trajectory but shifted relative to $t=0$.
        The histogram on the right-hand side of the plot shows the distribution of the number of infected ($\log_{10}$) at time \(t = 20\) from $5\times10^4$ stochastic simulations. 
        The histogram at the top is the empirical time-shifts distribution obtained by transforming the same simulations.
    }\label{fig:SIR_time_shift_and_trajectories}
\end{figure*}

As alluded to above, the form of \cref{eq:SIR_stochastic_sol,eq:SIR_mean_sol} are deliberate and suggest an elegant way to understand the effect of the early time stochasticity on the longer time mean dynamics. 
Equating the two solutions we see that the process \(I_b(t)\) can be approximated by \(I_d(t + \tau)\), where\begin{equation}\label{eq:time_shifts_SIR}
  \tau = \lambda^{-1} (\log W - \log \expect{W}).
\end{equation}
The two solutions are identical up to a random time-shifts of the initial conditions of the deterministic mean solution. 
The main panel of \cref{fig:SIR_time_shift_and_trajectories} illustrates this concept. The green lines show full stochastic realisations and the orange dashed lines indicate the corresponding shifted mean solutions {(see \citet{kendall:1966} for early prototype of this figure)}. 
The above analysis shows that the time-shift is a simple transformation of the random variable $W$.

\Cref{fig:SIR_time_shift_and_trajectories} summarises the main ideas of this section on the relationships between stochastic branching processes, deterministic approximations and time-shifts. 
The histogram at the top shows the time-shift distribution calculated from stochastic simulations run until $t=20$, which is enough time for $W(t)$ to have converged to $W$.
The histogram to the right shows the distribution of the state of the branching process ($I_b(t)$) at time $t = 20$, which can be seen to resemble the shape of the time-shift distribution but with some scaling. 
All simulations used to construct the histograms are conditioned on the event of non-extinction.

\subsection{Macroscopic dynamics}

The above discussion shows how the stochastic early time dynamics can be approximated by a deterministic solution for the mean subject to random initial conditions. As we saw in the discussion of the model dynamics, this noise persists on the macroscale as well. 
Theorem 1.1 of \citet{barbourEscapeBoundaryMarkov2015}, shows how the above time-shift analysis can be extended to the full non-linear deterministic approximation for the density process given in \cref{eq:SIR_det_approx}.
This states that given a CTMC with state vector \(\bs{X}(t)\) that is well approximated by a branching process near the initial condition and has a deterministic approximation, \(\bs{\zeta}(t)\), then the process \(\bs{\zeta}(t + \tau)\), with \(\tau\) given by \cref{eq:time_shifts_SIR}, also approximates \(\bs{X}(t)\).  
In our example here, $\bs{\zeta}(t) = N \bs{x}(t)$, where $\bs{x}(t) = (s(t),i(t))$.

Thus the time-shifts distribution has predictive power on the macroscale as well. For example, \cref{fig:SIR_time_delay}.B shows that we can predict the timing of peak infections as follows. 
Obtain a deterministic prediction for the timing of the peak number of infections, $t_p$, by either solving \cref{eq:SIR_det_approx} numerically or employing an analytic approximation \citep{turkyilmazoglu:2021}.
We can then determine the distribution on the peak timing from simply adding $t_p+\tau$. 

\section{Methods}
\label{sec:methods}

This section details the theoretical foundations for estimating the distribution of \(W\) and subsequently the distribution of the random time-shifts. 
We assume a branching process model is being used directly or has been derived from a CTMC model as was done for the SIR example in \cref{sec:SIR_example}. 

If a CTMC is the starting point, then the requirements for this analysis to be applicable are as follows:
The system has a natural scaling parameter $K$ (e.g. the total population or carrying capacity), such that one may consider how the dynamics scale in the limit $K\rightarrow \infty$. Typically this parameter is referred to as the system size and one talks about the dynamics in the deterministic limit \citep{blackStochasticFormulationEcological2012}.
For this limit to actually exist, it is required that the rates of events can be written in a density dependent form \citep{kurtzSolutionsOrdinaryDifferential1970,kurtzLimitTheoremsDiffusion1976,barbourDensityDependentMarkov1980}. 
This means that in the limit that $K\rightarrow \infty$, the density follows a set of ODEs (for example see \cref{eq:SIR_det_approx} for the SIR model, and \cref{app:innate_response_derivation} for the innate response model). Finally, we require that the rates of the CTMC are approximately linear near the initial condition so a BP approximation to the CTMC can be constructed \citep{barbourEscapeBoundaryMarkov2015,allenExtinctionThresholdsDeterministic2012,allenPrimerStochasticEpidemic2017}.
Many population models are naturally formulated such that all these conditions are simultaneously satisfied and hence the methodology is suitable (\citealp{barbourDensityDependentMarkov1980}; \citealp[Chapter~5.2]{schuster:2016}).

The three conditions above guarantee that we can approximate the early-time dynamics by a branching process, which can be \emph{matched} with a deterministic approximation in the same manner as in \cref{sec:SIR_BP,sec:random_time-shifts_SIR}. This matching can be done since the linearised (about the unstable equilibria) deterministic system produces an equivalent system of differential equations to the mean of the branching process. The resulting time-shifts distribution is guaranteed by Theorem~1.1 of \citet{barbourEscapeBoundaryMarkov2015} to accurately characterise the \emph{difference} between the true stochastic process and the limiting (large $K$) deterministic system.

The rest of this section is as follows: In \cref{sec:branching_processes} we detail the branching process theory required for defining \(W\) in the multivariate, continuous-time, case. 
\cref{sec:time_shifts} details our method for computing the LST of $W$ using a moment expansion and a suitably formulated embedded process. \cref{sec:lst_inversion} details the inversion of the LST to recover the distribution.
\cref{sec:moments} shows how the conditional moments required for the calculation of the LST can be found using a recursive approach. \cref{sec:discrete_app} shows how the method can be applied to discrete time models and \cref{sec:moment_matching} details a simpler, but approximate, approach for estimating the distribution of \(W\) by fitting a generalised gamma distribution to its first five moments.

\subsection{Branching process theory}\label{sec:branching_processes}

A branching process models the evolution of {the number of} particles, or agents, that each live for a particular lifetime at which point they die and reproduce according to predefined rules (\citealp[Chapter~V]{athreyaBranchingProcesses1972}; \citealp {dormanGardenBranchingProcesses2004}; \citealp[Chapter~1]{kimmelBranchingProcessesBiology2015}).
The particular property that makes a branching process amenable to analysis is that, once created, each individual is assumed to evolve independently of all others. 
{In a multi-type model each individual is assigned a type and each type is governed by different lifetimes and reproduction rules.}
We consider only Markovian branching processes where the lifetime for each type is assumed to be exponentially distributed, which we herein refer to as a  continuous-time multi-type branching process (CT-MBP) \citep{dormanGardenBranchingProcesses2004}.

Formulated mathematically, a CT-MBP \( \{ \bs{Z}(t), t\ge0 \}\) is defined on a state space \(\mathcal{S} \subseteq \mathbb{N}_0^m\) where  \(Z_i(t)\) is the number of individuals of type \(i = 1, \dots, m\), at time \(t\).
Each type lives for an exponentially distributed amount of time with mean \(1/a_i\), and upon death they create a number of offspring with \(p_{i}(\bs{k})\) the probability of individual \(i\) having \(\bs{k} = (k_1, \dots, k_m)\) offspring of each type. 
Note that throughout this work, unless specified otherwise, all vectors correspond to row vectors.
This information is conveniently summarised in the progeny generating functions 
\begin{equation} \label{eq:progeny_generating_function}
	f_i(\bs{s}) = \sum_{\bs{k}} p_i(\bs{k}) \prod_{j = 1}^{m} s_j^{k_j}, \quad \bs{s} \in [0, 1]^m.
\end{equation}
A CT-MBP is therefore fully specified by the rates \(a_i\) and the probabilities \(p_i(\bs{k})\).
Throughout this work we assume that the CT-MBP  is irreducible, that is for every pair of types $(i, j)$,

\begin{equation}
    \textrm{Pr}\, (Z_j(t) \ge 1 \,|\, \bs{Z}(0) = \bs{e}_i) > 0, \quad \textrm{for some } t\ge0,
\end{equation}

where \(\bs{e}_i\) is a standard basis vector in \(\mathbb{R}^m\).
Intuitively this means that starting with an individual of type $i$, there is a non-zero probability of eventually producing an individual of type $j$.

The dynamics of this model can be largely analysed through a matrix \(\Omega\) with elements (\citealp[Chapter~V]{athreyaBranchingProcesses1972}; \citealp{dormanGardenBranchingProcesses2004})
\begin{equation*}
	\Omega_{ij} = a_i \left( \eval{\dpd{f_i(\bs{s})}{s_j}}_{\bs{s} = 1} - \delta_{ij} \right),
\end{equation*}
where \(\delta_{ij} \) is the Kronecker delta.
In particular, defining \(\bs{\zeta}(t) := \mathds{E}[\bs{Z}(t)]\), we can characterise the mean behaviour of the branching process, which can be considered a deterministic approximation since \(\bs{\zeta}(t)\) satisfies the following system of ordinary differential equations \citep{dormanGardenBranchingProcesses2004}
\begin{equation}\label{eq:det_ode}
    \dod{}{t} \bs{\zeta}(t) = \bs{\zeta}(t) \Omega.
\end{equation}
Assuming the initial condition \(\bs{\zeta}(0) = \bs{z}_0\), the solution of \cref{eq:det_ode} is given by the matrix exponential
\begin{equation}\label{eq:det_solution}
    \bs{\zeta}(t) = \bs{z}_0 e^{\Omega t} .
\end{equation}
By Theorem 2.7 of \citet{senetaNonnegativeMatricesMarkov1981} and the Perron-Frobenius theorem (\citealp[Chapter V.7.4]{athreyaBranchingProcesses1972}; \citealp{barbourEscapeBoundaryMarkov2015}), in the limit as \(t\to \infty\) then \(e^{\Omega t} \approx e^{\lambda t} \bs{u}^T \bs{v} \) where \(\bs{u}^T\) (column vector) and \(\bs{v}\) (row vector) are the right and left eigenvectors of \(\Omega\), corresponding to the dominant eigenvalue \(\lambda\), normalised such that \(\bs{u} \cdot \bs{1} = 1\) and \(\bs{u} \cdot \bs{v} = 1\) (\citealp[Chapter~V]{athreyaBranchingProcesses1972}; \citealp[Chapter~VI]{harrisTheoryBranchingProcesses1964}).

We only consider the \textit{super-critical} regime where \(\lambda > 0\) as otherwise, with probability 1, the population will go extinct and hence not grow to be large on the macroscale.  
Hence, an approximate solution to \cref{eq:det_solution} valid at long times---but not so long that the BP approximation has broken down---is given by
\begin{equation}\label{eq:det_solution_approx}
    \bs{\zeta}(t) \approx e^{\lambda t} \boldsymbol{z}_0 \boldsymbol{u}^{T} \boldsymbol{v}.
\end{equation}

From these quantities a rescaled branching process can be defined,
\begin{equation}\label{eq:Wt}
	\bs{W}(t) = e^{-\lambda t} \bs{Z}(t),
\end{equation}
which is a non-negative Martingale with $\lim_{t \to \infty} \bs{W}(t) = W \bs{v}$
almost surely (\citealp{kestenLimitTheoremMultidimensional1966}; \citealp[Chapter V]{athreyaBranchingProcesses1972}).
This martingale is the main tool to understand how the mean dynamics differs from the  the stochastic realisations. 

For a long enough time such that \(\bs{W}(t)\) has converged, we can make the substitution  \(W \bs{v}\)  for \(\bs{W}(t)\) in \cref{eq:Wt} and rearrange to get \(\bs{Z}(t) \approx e^{\lambda t} W \bs{v}\) which can be expressed equivalently as \citep{barbourEscapeBoundaryMarkov2015}
\begin{equation*}
	\bs{Z}(t) \approx \exp{\left(\lambda \left( t +\lambda^{-1} \log W \right) \right) } \bs{v}.
\end{equation*}
The deterministic approximation for the mean, \cref{eq:det_solution}, can be written  as \(\bs{\zeta}(t) \approx e^{\lambda t}\mathds{E}[W] \bs{v}\) where \(\mathds{E}[W] = \bs{z}_0 \bs{u}^T\) and can be similarly expressed as
\begin{equation*}
	\bs{\zeta}(t) \approx \exp{\left(\lambda \left( t +\lambda^{-1} \log \expect{W}  \right) \right) } \bs{v}.
\end{equation*}

These two processes are identical up to the random time delay of \citep{barbourEscapeBoundaryMarkov2015}
\begin{equation}\label{eq:time_shifts_general}
	\tau =\lambda^{-1} (\log W - \log \expect{W}).
\end{equation}

The time-shift is a simple transformation of $W$ and thus the main part of our method is concerned with computing $W$ itself. 

The tractability of the later calculations relies on conditioning the process to start with a \emph{single} individual of a particular type. 
When the process starts from multiple individuals, i.e. a general initial condition $\boldsymbol{z}_0$, this is not a problem as we can exploit the independence of the agents to write
\begin{equation}\label{eq:branching_property}
	\bs{Z}(t) =  \sum_{i=1}^m \sum_{j = 1}^{z_{0, i}} \bs{Z}_i^{(j)}(t),
\end{equation}
where \(\bs{Z}_i^{(j)}\) are independent sub-processes that are each started from a single individual of type \(i\) for each of the \(j = 1, \dots, z_{0, i}\) initial individuals.
Defining $\bs{W}_i(t)$ as the random variable $\bs{W}(t)$ conditional on starting from a single individual of type $i$ and once more using Theorem 2, Chapter V.7 of \citet{athreyaBranchingProcesses1972} we have that
\begin{equation*}
	\lim_{t \to \infty} \bs{W}_i(t) = W_i \bs{v},
\end{equation*}
almost surely, and \(\mathds{E}[W_i] = u_i\).

Since the left eigenvector, $\bs{v}$, of the matrix $\Omega$ is the same regardless of initial condition we can multiply \cref{eq:branching_property} by \(e^{-\lambda t}\) and taking the limit as \(t \to \infty\) results in
\begin{equation}\label{eq:W_general_Z0}
	W  = \sum_{i=1}^m  \sum_{j = 1}^{z_{0, i}} W_i^{(j)}
\end{equation}
where the $W_i^{(j)}$, $j = 1, \dots, z_{0, i}$, are independent copies of $W_i$.

In order to compute the distribution of \(W\) we work with the Laplace-Stieltjes transforms (LSTs) of the $W_i$ defined as
\begin{equation}\label{eq:LST_i}
	\phi_i(\theta) = \mathds{E}[ e^{-\theta W_i}], \quad \theta \in \mathbb{C},
\end{equation}
and we define the vector \(\bs{\varphi}(\theta) = (\phi_1(\theta), \dots, \phi_m(\theta))\).
Since the random variables \(W_i^{(j)}\) are independent and identically distributed copies of the \(W_i\), the LST of \(W\) is simply
\begin{equation}\label{eq:LST_general_z0}
	\phi(\theta) = \prod_{i=1}^m \phi_i(\theta)^{z_{0, i}}.
\end{equation}

\subsection{Computation of the LST of $W_i$}\label{sec:time_shifts}

To compute the distribution of $W$ we will derive approximations for $\phi_i$, then \cref{eq:LST_general_z0} can be inverted using standard methods. 

Since \(\lambda > 0\), a Taylor series expansion of the term \(e^{-\theta W_i}\) about \(0\) in \cref{eq:LST_i} yields an approximation to the LST in terms of the first $n$ moments of $W_i$
\begin{equation}\label{eq:laplace_transform_approximation}
	\hat{\phi}_i(\theta) = \sum_{k = 0}^{n} \frac{(-\theta)^k}{k!} \mathds{E}[W_i^k].
\end{equation}
The calculation of the moments, $\mathds{E}[W_i^k]$, can be done by recursively solving sets of linear equations and is discussed in \cref{sec:moments}.
Simply evaluating \cref{eq:laplace_transform_approximation} at \(\theta \in \mathbb{C}\) will result in a poor approximation as \(\theta\) increases away from \(0\) due to the error term in the Taylor series \citep[Chapter 3.3]{hubbardVectorCalculusLinear1999}.

The error in the approximation can be determined through Lagrange's remainder theorem and the linearity property of expectation (see \cref{app:error_calculations_LT} for details).
Let $\mathcal{E}_i^{(n)}( \theta)$ denote the (absolute) error in \cref{eq:laplace_transform_approximation}, then this error is bounded above by 
\begin{equation}\label{eq:single_type_error}
    \mathcal{E}_i^{(n)}(\theta) \le \frac{\abs{\theta}^{n+1}}{(n+1)!} \mathds{E}[W_i^{n+1}].
\end{equation}
Since this holds for all $i = 1, \dots, m$, then we can simply ensure the largest LST error is below some (user specified) tolerance $\epsilon$ and the others will have error less than this by default.
Let 
\begin{equation*}
    \gamma = \max\left\{  \mathds{E}[W_i^{n+1}], i = 1, \dots, m\right\},
\end{equation*}
then 
\begin{equation*}
    \mathcal{E}_i^{(n)}(\theta) \le \frac{\abs{\theta}^{n+1}}{(n+1)!} \mathds{E}[W_i^{n+1}] \le \frac{\abs{\theta}^{n+1}}{(n+1)!} \gamma \le \epsilon,
\end{equation*}
and rearranging this error bound, we can define
\begin{equation}\label{eq:L_RHS_error}
    L(n, \epsilon) = \left(\frac{(n+1)!\epsilon}{\gamma}\right)^{1 / (n+1)}.
\end{equation}
Provided $\theta$ is in the region
\begin{equation}\label{eq:error_bound}
	\mathcal{A}_{n, \epsilon} = \left\{\theta : \abs{\theta} \le L(n, \epsilon)\right\},
\end{equation}
we simultaneously satisfy the error tolerance for all the LSTs. 
This provides a bound on the size of the open neighbourhood about \(0\) where the approximation has its error controlled to arbitrary levels of precision.

In order to extend the region where the approximation \cref{eq:laplace_transform_approximation} is accurate from \(\mathcal{A}_{n, \epsilon}\) to all of \(\mathbb{C}\), we consider the construction of an embedded discrete-time multi-type branching process (DT-MBP). 
Before we give the full details of the approach we provide insight into why this approach was the one taken.
The construction of the embedded DT-MBP means we can formulate another rescaled process (similar to \cref{eq:Wt}) which has the same limit, \(W \bs{v}\), as the original CT-MBP (\citealp{doobRegularityPropertiesCertain1940}; \citealp[Chapters~III.6 and V.7]{athreyaBranchingProcesses1972}). 
With a discrete-time BP the LSTs of the \(W_i\) can be shown to satisfy a simple functional equation that relates the progeny generating function (of the embedded process) and the LSTs.
This functional equation provides a method for \textit{shrinking} (we provide detail for what this means shortly) the value of \(\theta\) in the cases where \(\theta \notin \mathcal{A}_{n, \epsilon}\).
From the functional equation we can derive a simple recursive algorithm for evaluating the LSTs at \(\theta \in \mathbb{C}\).

The detailed realisation of this approach begins by constructing the embedded DT-MBP of \(\bs{Z}(t)\) which is defined, for some \(h > 0\), as \(\bs{Z}^{(h)}(n) = \bs{Z}(h n)\) with \(n \in \mathbb{N}_0\) \citep[Chapter~7.4]{modeMultitypeBranchingProcesses1971}.
{We provide results and some discussion for choosing the value of $h$ in \cref{sec:results}.}
The progeny generating functions for the embedded process can be calculated from the generating functions of the continuous-time process. 
{Once more we condition on the process starting with a single individual of type $i$ and define a rescaled version of the embedded chain
\begin{equation*}
	\bs{W}_i^{(h)}(n) = e^{-\lambda h n} \bs{Z}_i^{(h)}(n), \quad n \in \mathbb{N}_0, \quad i=1,\dots,m.
\end{equation*}
Crucially both the embedded process, \(\bs{W}_i^{(h)}(n)\), and original continuous time process, \(\bs{W}_i(t)\), converge to the same limit, \(W_i \bs{v}\) for all $i=1,\dots,m$ (\citealp{doobRegularityPropertiesCertain1940}; \citealp[Chapter~III]{athreyaBranchingProcesses1972}). }

{
Let $\tilde{f}_i(\boldsymbol{s})$ be the progeny generating function of the embedded process, $\bs{Z}^{(h)}(n)$, conditional on starting with an individual of type $i$ and let $\bs{\tilde{f}}(\boldsymbol{s}) = (\tilde{f}_1(\boldsymbol{s}), \dots, \tilde{f}_m(\boldsymbol{s}))$.
The progeny generating functions of the embedded process can be related to the progeny generating functions of the original process, $F_i(\bs{s},t)$,  where
\begin{equation*}
F_i(\bs{s},t) = \mathds{E}\left[\prod_{j=1}^{m} s_j^{Z_j(t)} \,\big\vert\, \bs{Z}(0) = \bs{e}_i\right], \quad i = 1, \dots, m.
\end{equation*}
These generating functions satisfy the system of differential equations (\citealp[Chapter~1.3]{allenStochasticPopulationEpidemic2015}; \citealp[Chapter~V]{athreyaBranchingProcesses1972}),
\begin{equation}\label{eq:odes_for_Z_pgfs}
	\dpd{F_i(\bs{s}, t)}{t}  = a_i \left( f_i(F_i(\bs{s}, t)) - F_i(\bs{s}, t) \right), \quad
 i=1,\dots,m,
\end{equation}
with initial conditions $F_i(\bs{s}, 0) = s_i$.
The progeny generating function of an individual of type $i$ in the embedded process is then \(\tilde{f}_i(\bs{s}) = F_i(\bs{s}, h)\).
Since the rescaled embedded process, $\bs{W}_i^{(h)}(t)$, and continuous-time processes, $\bs{W}_i(t)$, converge to the same limiting random variable, $W_i\bs{v}$, \(\bs{\varphi}(\theta)\) satisfies the following functional equation (\citealp[Chapter~V]{athreyaBranchingProcesses1972}; \citealp{harrisMathematicalModelsBranching1951}; \citealp[Chapter~1.8]{modeMultitypeBranchingProcesses1971}),
\begin{equation}\label{eq:discrete_functional}
	\bs{\varphi}(\theta) = \bs{\tilde{f}}\left( \bs{\varphi}\left( \theta e^{-\lambda h} \right) \right).
\end{equation}
}

This provides a way of accurately evaluating the approximation $\bs{\hat{\varphi}}(\theta)$ for all $\theta \in \mathbb{C}$. If $\theta \in \mathcal{A}_{n, \epsilon}$ then we simply evaluate \cref{eq:laplace_transform_approximation}.
If \(\theta \notin \mathcal{A}_{n, \epsilon}\) then we can recursively evaluate \cref{eq:discrete_functional} \(\kappa\) times until \(\theta e^{-\lambda h \kappa} \in \mathcal{A}_{n, \epsilon} \). 
This recursive calculation is equivalent to calculating
\begin{equation*}
    \bs{\varphi}(\theta) = \underbrace{\bs{\tilde{f}}\circ\bs{\tilde{f}}\circ\dots\circ\bs{\tilde{f}}}_{\kappa \textrm{ times}}\left( \bs{\varphi}\left( \theta e^{-\lambda\kappa h} \right) \right).
\end{equation*}
The value of \(\kappa\) can be chosen ahead of time for a particular value of \(L(n, \epsilon)\) (given by \cref{eq:L_RHS_error}) as
\begin{equation}\label{eq:choosing_kappa}
    \kappa \ge \frac{1}{\lambda h}\log{\left( \frac{\abs{\theta}}{L(n, \epsilon)} \right)}.
\end{equation}

The full procedure is listed below in \cref{alg:LST_inversion}. Note that the progeny generating functions for the embedded process, $\bs{\tilde{f}}(\bs{s})$, only needs to be calculated once for a given set of model parameters.  
The two hyperparameters are the number of moments used in the LST expansions, \(n\), and the size of the discrete time step, \(h\);
in \cref{sec:results} we will explore the choices of these parameters and their effect on the accuracy of the approximation and solve times. 

\begin{algorithm}[!htb]
    \caption{Computation of the LST of \(W\).}\label{alg:LST_inversion}
    \begin{algorithmic}[1]
        \Require $\bs{\tilde{f}}(\bs{s})$, $\theta$, $\lambda$, $h$, $\bs{z}_0$, $\epsilon$, $n$
        \State Compute the first $n + 1$ moments (see \cref{sec:moments})
        \State Compute $\mathcal{A}_{n, \epsilon}$ using the $(n+1)$th moments using \cref{eq:error_bound}
        \If{$\theta \in \mathcal{A}_{n, \epsilon}$}
            \State \(\bs{y} \gets \hat{\bs{\varphi}}(\theta)\) using \cref{eq:laplace_transform_approximation}
        \Else
            \State Calculate \(\kappa\) from \cref{eq:choosing_kappa}
            \State \(\bs{y} \gets \hat{\bs{\varphi}}(\theta e^{-\lambda \kappa h})\) using \cref{eq:laplace_transform_approximation}
            \For{\(i\) from \(1\) to \(\kappa\)} 
                \State \(\bs{y} \gets \bs{\tilde{f}}(\bs{y})\) using \cref{eq:discrete_functional}
            \EndFor
        \EndIf
        \State Return \(\hat{\phi}(\theta) \gets \displaystyle\prod_{i = 1}^{m} y_i^{z_{0, i}}\) 
    \end{algorithmic}
\end{algorithm}

\subsection{Distribution of $W$ using inversion}
\label{sec:lst_inversion}

We can use numerical inversion techniques to obtain the distribution of \(W\) from the LSTs. 
We refer to this method as the probability-estimation (PE) method. 
Inversion of $\phi(\theta) / \theta$ recovers the CDF, i.e.~$G_W(w) = \mathcal{L}^{-1}\left\{ \phi(\theta)/\theta \right\}(w)$ where $\mathcal{L}^{-1}$ is the inverse Laplace transform. 
This inversion can be carried out through a variety of methods, for example see \citet{abateIntroductionNumericalTransform2000, abateNumericalInversionLaplace1995} for an overview.
In this work we utilise the concentrated matrix exponential (CME) method \citep{horvathNumericalInverseLaplace2020}, with \(21\) terms. 
This method falls under the class of Abate-Whitt methods, out competes similar methods in terms of accuracy, and was robust throughout our testing. 
The CME approach, like most inversion methods, is valid only for values of $w > 0$. However, from analysis of the branching processes we know that there is a point mass at $w = 0$ corresponding to the probability of ultimate extinction, $q^{\star}$, \citep[Chapter~II.7]{harrisTheoryBranchingProcesses1964}, which is calculated below.
Hence we are able to add this in post-inversion and can express the CDF as
\begin{equation*}
\begin{aligned}
    G_W(w) &= \begin{cases}
        q^{\star}, & w = 0, \\ 
        \mathcal{L}^{-1}\left\{\hat{\phi}(\theta)/\theta \right\}(w), & w > 0.
    \end{cases}
\end{aligned}
\end{equation*}
Typically we will only be interested in the non-extinction case.
Defining the random variable \(W^\star := W \,\vert\, W > 0\), the probability density function (PDF) of this is
\begin{equation}
    g_{W^\star}(w) = \frac{1}{1 - q^\star} \dod{G_W(w)}{w}, \quad w > 0.
\end{equation}
The derivative of the CDF can be computed numerically or through automatic differentiation methods, which are more accurate \citep{baydinAutomaticDifferentiationMachine2018}.
In this work we utilise a specific version of automatic differentiation referred to as forward-mode automatic differentiation. 
This method is supported natively in Julia 
\citep{RevelsLubinPapamarkou2016}.

The probability $q^{\star}$ can be calculated as follows.
Let \(q_i\) be the probability of extinction conditioned on starting with a single individual of type \(i\) and define \(\bs{q} = (q_1, \dots, q_m)\).
The vector \(\bs{q}\) can be calculated by solving for the minimal non-negative solution of the system of equations {(\citealp[Chapter~II.7]{harrisTheoryBranchingProcesses1964}; \citealp[Chapter~7]{modeMultitypeBranchingProcesses1971})}
\begin{equation}\label{eq:odes_for_extinction}
	f_i(\bs{q}) = q_i, \quad i = 1, \dots, m.
\end{equation}
By the independence assumption of the individuals in the branching process the probability of ultimate extinction for a given initial condition is simply
\begin{equation*}
    q^\star = \prod_{i = 1}^{m} q_i^{z_{0,i}},
\end{equation*}
where \(z_{0, i}\) is the initial number of individuals of type \(i\).

\subsection{Calculating moments}\label{sec:moments}

In this section we outline the calculation of the moments $\expect{W_i^k}$ that are required in the Taylor expansion of the LST (\cref{eq:laplace_transform_approximation}).

The moment generating function (MGF) of $W_i$ is defined as
\begin{equation}\label{eq:mgfs}
	\xi_i(\theta) = \mathds{E}[ e^{\theta W_i}], \quad \theta \in \mathbb{R},
\end{equation}
and \(\bs{\Xi}(\theta) = (\xi_1(\theta), \dots, \xi_m(\theta)) \).
The \(i\)th MGF then satisfies the following functional equation (\citealp[Chapter~V]{athreyaBranchingProcesses1972})
\begin{equation}\label{eq:functional_continuous}
	\xi_{i}(\theta) = \int_{0}^{\infty} f_i \left( \bs{\Xi}(\theta e^{-\lambda y}) \right) a_i e^{-a_i y} \dif y,
\end{equation}
where \(f_i(\bs{s})\) is the progeny generating function and \(a_i\) is the rate parameter of the exponential lifetime distribution for individuals of type \(i\) respectively. 

We note that throughout this section, unless otherwise specified, we use the notation $\xi_i^{(n)}(x)$ to denote the $n$th derivative of $\xi_i(\theta)$ with respect to $\theta$ evaluated at $x$, and 
let $\boldsymbol{\Xi}^{(n)}(x) = (\xi_1^{(n)}(x), \dots, \xi_m^{(n)}(x))$.

The \(n\)th derivative of the \(i\)th MGF yields the \(n\)th moment of \(W_i\), $\xi_i^{(n)}(0) = \mathds{E}[W_i^n]$ and hence all the moments 
can be obtained by differentiating \cref{eq:functional_continuous} \(n\) times, {for each $i$}, and evaluating the result at \(\theta = 0\) \citep{bellmanAgeDependentBinaryBranching1952},
\begin{equation}\label{eq:interchanging_int_deriv}
	\xi_i^{(n)}(0) = \int_{0}^{\infty} \eval{\dpd[n]{}{\theta} f_i \left( \bs{\Xi}(\theta e^{-\lambda y}) \right)}_{\theta = 0} a_i e^{-a_i y} \dif y.
\end{equation}
{Upon first consideration it would appear simpler to differentiate \cref{eq:discrete_functional} to obtain the moments. 
However evaluation of the progeny generating functions for the embedded process requires a system of ODEs to be solved numerically (see \cref{sec:discrete_app}), which is not easily handled.}
The process of repeated differentiation of the progeny generating functions is complicated in general,
{however frequently occurring reproduction rules for a branching processes (i.e. linear or quadratic progeny generating functions) can be solved for and yield a system of linear equations that when solved gives the moments for each $W_i$. 
}
This enables the calculations to be automated and examples of this for common use cases will be provided here.
{This approach is also exact in the sense that it is not influenced by the accuracy of the numerical solvers for obtaining the ODE solutions that are needed for evaluating for the progeny generating functions.}

{
In this work we consider progeny generating functions of the form 
\begin{equation}\label{eq:general_pgf}
    f_i(\boldsymbol{s}) = \frac{\nu_i}{a_i} + \sum_{\substack{j = 1 \\ j \neq i}}^{m}\frac{\alpha_{ij}}{a_i} s_j + \sum_{k = 1}^{m} \sum_{l = k}^{m} \frac{\beta_{ikl}}{a_i} s_k s_l,
\end{equation}
where the rate parameter for the lifetime distribution is given by 
\begin{equation*}
    a_i = \nu_i + \sum_{\substack{j = 1 \\ j \neq i}}^{m} \alpha_{ij} + \sum_{k = 1}^{m} \sum_{l = k}^{m} \beta_{ikl}.
\end{equation*}
The parameter $\nu_i$ relates to the rate of type $i$ dying without producing any offspring; the parameters $\alpha_{ij}$ correspond to linear branching dynamics (i.e. type $i$ dying and generating a type $j$), and the $\beta_{ikl}$ correspond to quadratic branching (i.e. type $i$ splitting into two other types $k$ and $l$). 
The summation involving the $\alpha_{ij}$ is over values $\{1, \dots, m\} \setminus \{i\}$ as type $i$ must become another type in this case. 
This is not a restriction in the quadratic branching case (i.e. the summation including the $\beta_{ikl}$) however we do assume an ordering $l \ge k$ on the indices, which assures there is no double counting (i.e. no contribution for $s_k s_l$ and $s_l s_k$ as these are treated the same and hence $\beta_{ikl} = 0$ for $k > l$).

Substituting \cref{eq:general_pgf} into \cref{eq:functional_continuous} and noting that as the sums are finite we can swap the order of summation and integration, the functional equation takes the form 
\begin{equation}\label{eq:functional_continuous_general}
    \xi_i(\theta) = I_1(\theta) + I_2(\theta) + I_3(\theta)
\end{equation}
where 
\begin{align}
    I_1(\theta) &= \nu_i \int_{0}^{\infty}  e^{-a_i y} \dif y, \label{eq:functional_continuous_general_1} \\ 
    I_2(\theta) &= \sum_{\substack{j = 1 \\ j \neq i}}^{m} \alpha_{ij} \int_{0}^{\infty}  \xi_j(\theta e^{-\lambda y}) e^{-a_i y} \dif y, \label{eq:functional_continuous_general_2} \\ 
    I_3(\theta) &= \sum_{k = 1}^{m} \sum_{l = k}^{m} \beta_{ikl} \int_{0}^{\infty}  \xi_k(\theta e^{-\lambda y}) \xi_l(\theta e^{-\lambda y}) e^{-a_i y} \dif y \label{eq:functional_continuous_general_3}.
\end{align}
We can therefore differentiate \cref{eq:functional_continuous_general} $n$ times and evaluate at $\theta = 0$ by considering the terms in \cref{eq:functional_continuous_general_1,eq:functional_continuous_general_2,eq:functional_continuous_general_3} individually. 
Carrying this out and substituting the results we arrive at the equation

\begin{equation}\label{eq:general_equation_scalar_not_simple}
\begin{aligned}
    \xi_i^{(n)}(0) =& \sum_{\substack{j = 1 \\ j \neq i}}^{m} \frac{\alpha_{ij}}{a_i + n \lambda} \xi_j^{(n)}(0) \\ &+ \sum_{k = 1}^{m} \sum_{l = k}^{m} \frac{\beta_{ikl}}{n\lambda + a_i}\sum_{r = 0}^{n} \binom{n}{r} \xi_k^{(r)}(0) \xi_l^{(n - r)}(0).
\end{aligned}
\end{equation}
With the initial condition $\xi_i^{(1)}(0) = \mathds{E}[W_i] = u_i$ (see \cref{sec:branching_processes}), we can formulate a recursive system of linear equations (in the moments) by isolating all the terms involving the $n$th moment on the left-hand side of \cref{eq:general_equation_scalar_not_simple}. 
The recursive equation is then given by
\begin{multline}\label{eq:general_equation_scalar}
    \xi_i^{(n)}(0) - \sum_{\substack{j = 1 \\ j \neq i}}^{m} \frac{\alpha_{ij}}{a_i + n \lambda} \xi_j^{(n)}(0) - \sum_{k = 1}^{m} \sum_{l = k}^{m} \frac{\beta_{ikl}}{n\lambda + a_i}\left( \xi_k^{(n)}(0) + \xi_l^{(n)}(0) \right) \\ 
    = \sum_{k = 1}^{m} \sum_{l = k}^{m} \frac{\beta_{ikl}}{n\lambda + a_i} \sum_{r = 1}^{n-1} \binom{n}{r} \xi_k^{(r)}(0) \xi_l^{(n - r)}(0), \quad n \ge 2.
\end{multline}
Defining the constants
\begin{equation*}
\begin{aligned}
    \tilde{\alpha}_{ij}^{(n)} &= \frac{\alpha_{ij}}{a_i + n \lambda}, \\ 
    \tilde{\beta}_{ikl}^{(n)} &= \frac{\beta_{ikl}}{n\lambda + a_i}, \\ 
    d_i^{(n)} &= \sum_{k = 1}^{m} \sum_{l = k}^{m} \tilde{\beta}_{ikl}^{(n)}\sum_{r = 1}^{n-1} \binom{n}{r} \xi_k^{(r)}(0) \xi_l^{(n - r)}(0),
\end{aligned}
\end{equation*}
and noting that $\xi_i^{(n)}(0) = \boldsymbol{e}_i \boldsymbol{\Xi}^{(n)}(0)^{T}$ we can rewrite \cref{eq:general_equation_scalar} in vector notation
\begin{equation}\label{eq:general_equation}
    \left( \boldsymbol{e}_i - \sum_{\substack{j = 1 \\ j \neq i}}^{m} \tilde{\alpha}_{ij}^{(n)} \boldsymbol{e}_j - \sum_{k = 1}^{m}\sum_{l = k}^{m} \tilde{\beta}_{ikl}^{(n)} \left( \boldsymbol{e}_k + \boldsymbol{e}_l \right) \right) \boldsymbol{\Xi}^{(n)}(0)^{T} = d_i^{(n)}, \quad n \ge 2,
\end{equation}
subject to the initial condition $\boldsymbol{\Xi}^{(1)}(0) = \boldsymbol{u}$.

\Cref{eq:general_equation} holds for agents of type $i = 1, \dots, m$ and so we can formulate a matrix \(C^{(n)}\) and a vector \(\bs{d}^{(n)}\), that constitute a linear system that can be solved recursively to get the \(n\)th moments,
\begin{equation*}
    C^{(n)} \bs{\Xi}^{(n)}(0)^{T} = \bs{d}^{(n)T}, \textrm{ for } n \ge 2.
\end{equation*}
Row $i$ of $C^{(n)}$ corresponds to a linear equation derived from the $i$th progeny generating function. 
An example of formulating this system of equations is given for the SEIR epidemic model in \cref{sec:SEIR_results}. 
}

\subsection{Application to discrete-time models}
\label{sec:discrete_app}

The method developed so far has been for continuous-time processes, but discrete-time processes can also be handled with some simplifications to the method that we briefly outline in this section.
A discrete-time multi-type branching process (DT-MBP) \(\bs{Z}(t)\) is specified similarly to the CT-MBP but with \(t \in\mathbb{N}_0\).
The progeny generating functions take the same form as \cref{eq:progeny_generating_function} but individuals have non-random unit lifetimes. 
The behaviour of the system is studied through the mean offspring matrix \(\mathcal{M}\) with elements
\begin{equation*}
    \mathcal{M}_{ij} = \eval{\dpd{f_i(\bs{s})}{s_j}}_{\bs{s} = 1} \quad i, j = 1,\dots, n. 
\end{equation*}
The dominant eigenvalue of this matrix is denoted by \(\rho\) and with it we define the analogous form of \cref{eq:Wt}, \(\bs{W}(t) = \bs{Z}(t) \rho^{-t}\) \citep[Chapter~V]{athreyaBranchingProcesses1972}. 
Note that \(\rho^{-t} = e^{-t\log\rho}\) and letting \(\lambda = \log\rho\) we have \(\bs{W}(t) = \bs{Z}(t) e^{-\lambda t}\) as in the continuous-time case (\cref{eq:Wt}). 
The rescaled process \(\bs{W}(t)\) approaches the limit \(W \bs{v}\) where \(\bs{u}^T\) and \(\bs{v}\) now correspond to the right and left eigenvectors of \(\mathcal{M}\), normalised such that \(\bs{u} \cdot \bs{1} = 1\) and \(\bs{u} \cdot \bs{v} = 1\).

All the remaining constructions from the previous sections apply, but with the simplification that the conditional LSTs of \(W\) directly satisfy (\citealp[Chapter~V]{athreyaBranchingProcesses1972}; \citealp[Chapter~1.8]{modeMultitypeBranchingProcesses1971})
\begin{equation}\label{eq:functional_discrete_time}
    \bs{\varphi}(\theta) = \bs{f}\left( \bs{\varphi}\left( \theta \rho^{-1} \right) \right). 
\end{equation}
\cref{alg:LST_inversion} can be used to compute the LSTs by setting \(\bs{\tilde{f}}(\bs{s}) = \bs{f}(\bs{s})\), \(\lambda = \log \rho\) and \(h = 1\).
For a discrete-time model, the process of deriving the moments, given a model specification, is dramatically simplified:
firstly, as we can directly differentiate \cref{eq:functional_discrete_time} (as opposed to differentiating \cref{eq:functional_continuous} as in the continuous-time case) to obtain the moments by replacing $\bs{\varphi}(\theta)$ with the MGFs, $\bs{\Xi}(\theta)$. 
Secondly, we do not need to extend the neighbourhood for evaluating the LST through solving ODEs, as the progeny generating functions are explicitly given.

\subsection{Moment matching}\label{sec:moment_matching}

Here we outline a second approach to calculating the distribution of \(W\) that is based on moment matching with a parametric distribution.
This approach is an approximation, but is quicker than the PE method as we do not require construction of the LST and its subsequent inversion. 
Furthermore, this method results in analytical distributions which can be more easily sampled.

First, recall from \cref{eq:W_general_Z0} that $W$ can be written as the sum of independent random variables $W^{(j)}_i$ and so the $k$th moment of $W$ is given by 
\begin{equation*}
    \expect{W^k} = \mathds{E}\left[ \left(  \sum_{i = 1}^{m} \sum_{j = 1}^{z_{0, i}} W_{i}^{(j)} \right)^{k}\right].
\end{equation*}
For ease of notation, define $N = \sum_{i = 1}^{m} z_{0, i}$ independent random variables $U_1 = W_1^{(1)}, U_2 = W_1^{(2)}, \dots, U_{z_{0, 1}} = W_1^{(z_{0, 1})}, \dots, U_{N} = W_m^{(z_{0, m})}$, then the moments of $W$ can be written as
\begin{align}\label{eq:W_combination_mm}
    \expect{W^k} &= \mathds{E}\left[ \left(  \sum_{n = 1}^{N} U_n \right)^{k}\right] = \sum_{\boldsymbol{l} \in \mathcal{B}_k} \binom{k}{l_1, l_2, \dots, l_N} \prod_{n = 1}^{N} \mathds{E}\left[ U_n^{l_n} \right],
\end{align}
which follows from the multinomial theorem and the linearity of expectation over finite sums. 
The set $\mathcal{B}_k$ is defined as 
\begin{equation*}
    \mathcal{B}_k = \left\{ \boldsymbol{l} : \sum_{n = 1}^{N} l_n = k, l_n \in \{0, 1, \dots, k\} \right\},
\end{equation*}
which are the integer partitions of $k$. 
The expectations appearing in \cref{eq:W_combination_mm}, $\mathds{E}[U_n^{l_n}]$, are simply the moments of the $W_i^{(j)}$'s that are calculated in \cref{sec:moments}.

Next, recall from \cref{sec:lst_inversion} that the distribution of $W$ can be expressed as a mixture of a point mass at $w=0$ and a continuous part for $w>0$, which is denoted $W^\star$. We can therefore approximate $W$ by fitting a parametric distribution to $W^\star$, adding the point mass, and re-normalising appropriately. 
The family of distributions we fit to is chosen by considering the known properties of \(W^\star\).
The distribution of \(W^\star\) is strictly non-negative and absolutely continuous \citep[Chapter~V]{athreyaBranchingProcesses1972}.
Additionally, we assume that the distribution of the sample paths of the branching processes at time \(t\) are unimodal and potentially heavy-tailed away from \(0\), which implies similar characteristics for {\(\bs{W}^{\star}(t)\) and hence \(W^{\star}\).
This suggests that suitable distributions would likely be from the exponential family and as such we consider fitting a generalised gamma distribution 
by using a moment matching method (MM).
This distribution has the Weibull, exponential and gamma distributions as special cases and in testing appeared to be well fitting. 

{
Suppose \(W^{\star} \sim \textrm{GG}(\beta, \alpha_1, \alpha_2)\), then the PDF is given by 
\begin{equation*}
	g_{W^\star}(w) = \frac{\alpha_2}{ \beta^{\alpha_1}\Gamma(\alpha_1/\alpha_2)}  w^{\alpha_1 - 1} \exp{\left(-\left(\frac{w}{\beta}\right)^{\alpha_2}\right)}, 
\end{equation*}
with the \(k\)th moment given by
\begin{equation*}
	M_k(\beta, \alpha_1, \alpha_2) = \beta^{ k} \frac{\Gamma((\alpha_1 + k)/\alpha_2)}{\Gamma(\alpha_1 / \alpha_2)}.
\end{equation*}
We determine the parameters of this distribution, $(\beta, \alpha_1, \alpha_2)$, by minimising the difference between its first five moments and those of \(W^{\star}\) as calculated by our method, where 
$\expect{W^{\star k}} = (1 - q^\star)^{-1} \expect{W^{k}}.$

Since the moments grow very large with increasing $k$, we standardise them to ensure that each are assigned approximately equal importance when fitting \citep[Chapter~8]{bishopNeuralNetworksPattern1996}. 
This ensures that the optimisation routine does not prioritise the higher order moments that would otherwise excessively contribute to a naive loss function.
To determine the appropriate scaling we express the moments of $W^{\star}$ as $\expect{W^{\star k}} = c_k \times 10^{\eta_k}$ 
for some $c_k \in [0, 10)$ and some $\eta_k$.
From this we can define the loss function as the sum of squares
\begin{equation}\label{eq:loss_func}
	L(\beta, \alpha_1, \alpha_2) = \sum_{k=1}^{5} \left(\frac{ \expect{W^{{\star} k}} - M_k(\beta, \alpha_1, \alpha_2)  }{10^{\eta_k}}\right)^2.
\end{equation} 
This can be minimised numerically using standard methods to estimate the parameters.

Samples can be drawn from the GG distribution using the inverse CDF method.
Additionally, the approximation to the time-shift distribution conditional on non-extinction, denoted $\tau^\star$, can be derived through the CDF method and has PDF given by
\begin{equation}
    g_{\tau^\star}(\tau) = \frac{\alpha_2\lambda\mu_W^{\alpha_1}}{\beta^{\alpha_1} \Gamma\left(\frac{\alpha_1}{\alpha_2}\right)} \exp{\left( -\left( \frac{\mu_W e^{\lambda \tau}}{\beta} \right)^{\alpha_2} + \alpha_1\lambda \tau \right)}.
\end{equation}

\section{Results}\label{sec:results}

In \cref{sec:methods} we provided the details of two approaches for computing the distribution of $W$ and hence the time-shift itself. 
The first method, which we refer to as the PE method, relies on approximating the LST of $W$ and inverting it (detailed in \cref{sec:lst_inversion}) and the second, called the MM  method, relies on matching the first five moments (detailed \cref{sec:moment_matching}). This section focuses on how these methods perform on three models of increasing complexity.

We first demonstrate how the methods perform on the SIR model described in \cref{sec:SIR_example}, since this is a special case where the distribution of $W$ is known analytically. 
In \cref{sec:SEIR_results} we apply the methods to an SEIR model, which is a simple extension of the SIR example.
\cref{sec:results_hyper_params} explores the effect of the hyper-parameters (the number of moments in the expansion and the step size for the embedded process \(h\)) on the accuracy of the resulting distributions as well as the computation time. 
Following this, in \cref{sec:results_system_size} we assess the effect of the total population size, on the time-shift distributions which provides some insight to when this method is suitable.
Finally, in \cref{sec:results_initial_conditions} we explore more complex initial conditions and how this influences the shape and location of the resulting time-shift distributions. 
In \cref{sec:innate_response_results} we explore a more complicated model which demonstrates how the method can be used to approximate the macroscopic dynamics of a more complex 6-state CTMC model. 

{
\subsection{SIR model}
\label{sec:SIR_results}

The first example is the SIR model (see \cref{sec:SIR_example} for formulation) which serves as validation of the method for a one-dimensional situation where the distribution of $W$, and hence the distribution of $\tau$, is known analytically.
This follows as the early time approximation of the SIR model is a one-dimensional, linear, birth-death process \citep{harrisMathematicalModelsBranching1951}. 
We give the main results here; a full derivation is given in \cref{app:SIR_LST}. 
For this model the LST of $W$ is given by 
\begin{equation}\label{eq:LST_SIR}
    \phi(\theta) = \frac{\gamma}{\beta} + \frac{\beta - \gamma}{\beta} \left( 1 + \frac{\beta \theta}{\beta - \gamma} \right)^{-1}.
\end{equation}
Inversion of $\phi(\theta) / \theta$ gives the CDF of $W$, 
\begin{equation}\label{eq:CDF_SIR}
    G_{W}(w) = q + (1 - q) \left(1 - e^{-(1-q) w}\right), \quad w \ge 0,
\end{equation}
where $q = \gamma / \beta$. 
The point mass of size $q$ at $w=0$ corresponds to extinction of the process. 
Conditional on the event of non-extinction ($w>0$), the distribution is exponential with rate \(1 - q\).

For the results in this section we fix the initial condition at \(\bs{X}(0) = (N-1, 1)\) where \(N = 10^6\) and the parameters at \((\beta, \gamma) = (0.95, 0.5)\) (as they were in \cref{sec:SIR_example}).
We choose the two hyper-parameters, the time step, $h = 0.1$, and number of terms in the moment expansion, $n = 30$.
The choices of these parameters and the effect of these choices are assessed in the following section (\cref{sec:SEIR_results}).

\Cref{fig:SIR_lst_cdf_comparison} shows the LST and CDF computed using our methods (green dots) against the true values (black solid line) as given by \cref{eq:LST_SIR} and \cref{eq:CDF_SIR}, respectively. 
Note that we only show the PE method here as the MM method is visually indistinguishable. 
We see strong agreement between the exact quantities and their counterpart estimated with our methods where the shape of the LST is appropriately captured.

\begin{figure*}[!htb]
    \centering
    \includegraphics{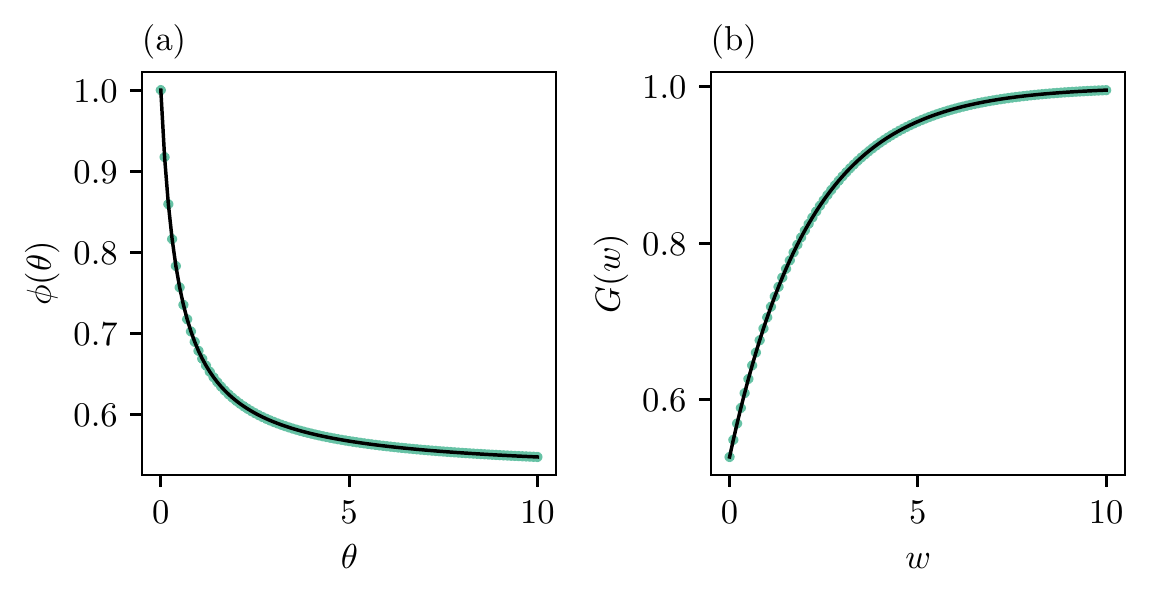}
    \caption{
		Panel (a): Comparison of the exact LST (\cref{eq:LST_SIR}) (black solid line) and our approximation from the two methods (green dots). 
        Panel (b): Comparison of the CDF (\cref{eq:CDF_SIR}) (black solid line) and our approximation from the two methods (green dots). 
        Note that in both panels, the results for the two methods (PE and MM) are visually indistinguishable so we show only one (PE) for clarity.
        Numerical results for both methods are provided in \cref{tbl:sir_results}. 
	}
	\label{fig:SIR_lst_cdf_comparison}
\end{figure*}

To assess the accuracy of the different methods more quantitatively we consider the simple measure of the L1-norm \citep[Chapter 13]{pajankarHandsonMachineLearning2022} between the CDF values computed under \cref{eq:CDF_SIR} and our methods over the interval $[0, 10]$ in steps of $0.1$. 
This can be considered as an average of the error over the interval.
Alongside this we also report the maximal error over the interval. 

\Cref{tbl:sir_results} shows the results of these two values. 
We see that both methods produce low average errors based on the L1-norm. 
While the maximum error occurs for the PE method (on the order of $10^{-5}$) which is 5 orders of magnitude larger than the error for the MM method, it should be noted that the error corresponds to a difference in the 4th decimal place. 
This will be largely insignificant for practical use cases and as seen in \cref{fig:SIR_lst_cdf_comparison}, the methods can reliably approximate the LST and CDF.

\begin{table}[!htb]
	\centering
	\begin{tabular}{@{}ccc@{}}
		\toprule
		Method         & L1-norm                     & Max error \\ \midrule
        PE & $9.978\times10^{-5}$ & $1.478\times10^{-4}$ \\ 
        MM & $9.339 \times10^{-10}$ & $1.235\times10^{-9}$
		\\
		\bottomrule
	\end{tabular}
	\caption{
        Error between the two estimation methods and the true CDF for the SIR example. 
        Results are computed over the interval $[0, 10]$ in steps of $0.1$.
    }
	\label{tbl:sir_results}
\end{table}
}

\subsection{SEIR model}
\label{sec:SEIR_results}

The next example is the canonical extension to the SIR model that incorporates a latent (or exposure) state where individuals are infected but not yet infectious \citep[Chapter~2.5]{keelingModelingInfectiousDiseases2008}.
Assuming a fixed population of size \(N\), we can formulate the SEIR model as a CTMC with state vector \(\bs{X}(t) = (S(t), E(t), I(t))\).
The model is governed by the parameters \((\beta, \sigma, \gamma)\) where \(\beta\) is the effective transmission parameter, \(\sigma\) is the rate of transitioning from \(E\) to \(I\), and \(\gamma\) is the rate of transitioning from \(I\) to \(R\).
When \(\bs{X}(0) \approx (N, 0, 0)\) we can approximate the CTMC by a 2-type CT-MBP, \(\bs{Z}(t) = (E(t), I(t))\).
Rates of the CTMC and CT-MBP approximation are given in \cref{tbl:seir_model_rates}.
Note that we overload notation here and use the same variables ($S, E$ and $I$) for both the population numbers and states.

\begin{table}[!htb]
	\centering
	\begin{tabular}{@{}cccc@{}}
		\toprule
		\(\Delta \bs{X}\)         & CTMC Rate                    & \(\Delta \bs{Z}\)         & Linearised (BP) Rate \\ \midrule
		\((S, E) \to (S-1, E+1)\) & \(\dfrac{\beta I S}{N - 1}\) & \(E \to E+1\)             & \(\beta I\)          \\
		\((E, I) \to (E-1, I+1)\) & \(\sigma E\)                 & \((E, I) \to (E-1, I+1)\) & \(\sigma E\)         \\
		\(I \to I-1\)             & \(\gamma I\)                 & \(I \to I-1\)             & \(\gamma I\)         \\
		\bottomrule
	\end{tabular}
	\caption{Change in state and rates for the CTMC SEIR model and the BP approximation.}
	\label{tbl:seir_model_rates}
\end{table}

We use the natural ordering of the CT-MBP state vector to define the mapping between the types so that \(E = \textrm{type }1\) and \(I = \textrm{type }2\) individuals.
The lifetimes of individuals are exponentially distributed with rates \(a_1 = \sigma\) and \(a_2 = \beta + \gamma\).
An individual of type 1 produces a single offspring of type 2 after their lifetime with probability 1.
Individuals of type 2 either produce no offspring with probability \(\gamma / a_2\) or produce {one offspring of type 1 and type 2} with probability \(\beta / a_2\).
Thus, the PGFs for the CT-MBP are
\begin{align}
	f_1(\bs{s}) & = \frac{\sigma}{a_1} s_2 \label{eq:eg_linear},                                            \\
	f_2(\bs{s}) & = \frac{\gamma}{a_2} + \frac{\beta}{a_2}\beta s_1 s_2 \label{eq:eg_quadratic}.
 \end{align}

Next we formulate the recursive system of equations needed for computing the moments through the method outlined in \cref{sec:moments}.
We can easily extract the rate constants from \cref{eq:eg_linear} and \cref{eq:eg_quadratic} using the subscripts of the progeny generating function and the subscripts of the elements of $\boldsymbol{s}$ appearing in the right-hand side.
In \cref{eq:eg_linear}, $i = 1$ (from the left-hand side) and $j = 2$ (from the right-hand side) so $\alpha_{12} = \sigma$. 
Similarly, in \cref{eq:eg_quadratic}, $i = 2$, $k = 1$ and $l = 2$, so $\beta_{212} = \beta$ for the second equation. 
Hence the only non-zero parameters appearing in the rows of $C^{(n)}$ arising from \cref{eq:general_equation} are
\begin{equation*}
    \tilde{\alpha}_{12}^{(n)} = \frac{\sigma}{\sigma + n \lambda}, \quad \tilde{\beta}_{212}^{(n)} = \frac{\beta}{\beta + \gamma + n \lambda},
\end{equation*}
noting that the leading subscript denotes which row of $C^{(n)}$ the parameters correspond to. 
The growth rate, $\lambda$, for the SEIR model is given explicitly by \citep{maEstimatingEpidemicExponential2020a} 
\begin{equation*}
    \lambda = \frac{-(\sigma + \gamma) + \sqrt{(\sigma-\gamma)^2 + 4 \sigma \beta}}{2}.
\end{equation*}
With the constants determined, the system of linear equations for the moments is given succinctly as
\begin{equation*}
    \begin{bmatrix}
        1 & -\tilde{\alpha}_{12}^{(n)} \\ 
        -\tilde{\beta}_{212}^{(n)} &\quad 1 - \tilde{\beta}_{212}^{(n)} 
    \end{bmatrix} \begin{bmatrix}
        \xi_1^{(n)}(0) \\ 
        \xi_2^{(n)}(0)
    \end{bmatrix} = \begin{bmatrix}
        0 \\ 
        \tilde{\beta}_{212} \displaystyle{\sum_{r = 1}^{n-1}} \binom{n}{r} \xi_j^{(r)}(0)\xi_k^{(n - r)}(0) 
    \end{bmatrix}, \quad n \ge 2, 
\end{equation*}
with $(\xi_1^{(1)}(0), \xi_2^{(1)}(0)) = (u_1, u_2)$, which is the normalised eigenvector corresponding to the eigenvalue $\lambda$ as outlined in \cref{sec:branching_processes}. 

\subsubsection{Hyper-parameter sensitivity}\label{sec:results_hyper_params}

The two free hyper-parameters of the algorithm are: 
the number of terms in the moment expansion, $n$, and 
the choice of the time step, \(h\), for the construction of the embedded process.
For the simulations in this section and the following sections (unless specified otherwise) we fix the initial condition at \(\bs{X}(0) = (N-1, 1, 0)\) where \(N = 10^6\) and the parameters at \((\beta, \sigma, \gamma) = (0.56, 0.5, 0.33)\). 
Different parameter choices (satisfying \(\lambda > 0\)) were also explored and the results were consistent with those presented for this parameter combination\footnote{In fact, these parameters correspond to choices of a basic reproduction number, $R_0 = \beta / \gamma = 1.7$, average latent period, $\sigma^{-1} = 2$, and average infectious period, $\gamma^{-1} = 3$.}.

Exact stochastic simulation \citep{gillespieExactStochasticSimulation1977} is utilised throughout this section to estimate the empirical distribution of the time-shifts by computing the difference in the time for the number of infected to reach a threshold of \(0.05 N\) under the stochastic simulation and deterministic approximation respectively. This threshold was chosen based on the time taken for an ensemble of stochastic simulations to have all reached their exponential growth phase, measured by comparing when the growth rate of a simulation was consistent with the deterministic model.
All simulations will be conditioned on the event of non-extinction and we take the error tolerance to be \(\epsilon = 1\times 10^{-6}\) (used in \cref{eq:error_bound}).

\Cref{fig:effect_of_n} shows the distribution of the time-shifts for different choices of the number of moments, $n$. 
We see that fewer terms ($n = 3$) in the expansion results in poor estimation of the PDF.
Oscillations appear in the PDF estimated through our method and these become increasingly large in the right tail of the distribution. 
These oscillations are clear in the $n = 3$ case and also exist in the $n = 10$ and $n = 15$ cases but in a less obvious way. 
There appears to be minimal difference in the choice of \(n\) once \(n > 15\) and at $n = 30$ the expansion is highly accurate.
In what follows, unless stated otherwise, we assume $n = 30$ moments are used in the expansion.

\begin{figure*}[!htb]
    \includegraphics{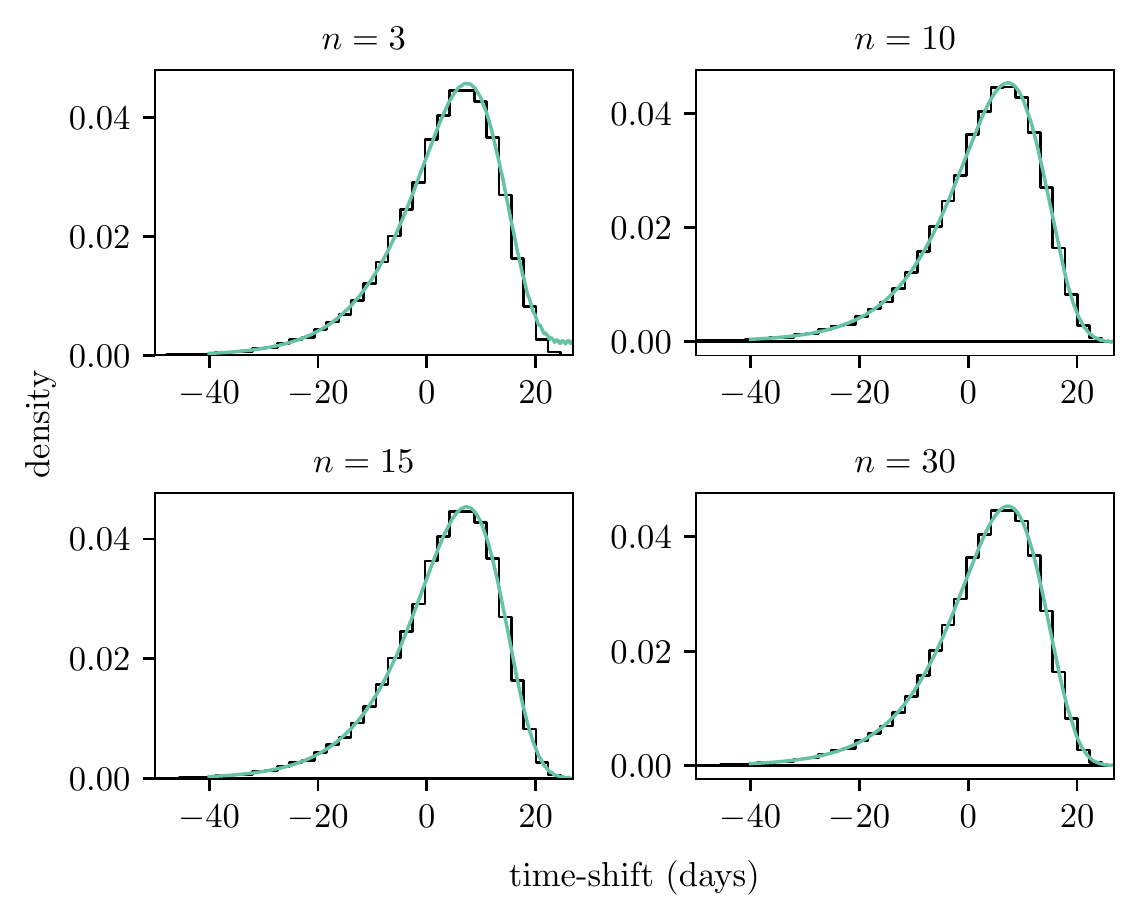}

    \caption{
        Effect of the number of terms, $n$, in the moment expansion. 
        Each panel shows the PDF for the random time-shifts with a varying number of terms where the value of $n$ is given in the title. 
        The black histogram shows the empirical PDF and the coloured lines show the estimated density from the PE method. 
    }\label{fig:effect_of_n}
\end{figure*}

\Cref{fig:effect_of_h} shows the time-shift distributions for \(4\) different choices for the time-step, \(h\), of the embedded process. 
The choices reflect some typical values for a model of this size and temporal resolution, but the computed distributions are visually indistinguishable, hence the method appears to be insensitive to \(h\). 
The estimated PDFs do not exhibit any deviations from the empirical time-shift distribution which suggests that the choice of \(n\) is more critical to the accurate estimation of the time-shift distributions. 

\begin{figure*}[!htb]
    \centering
    \includegraphics{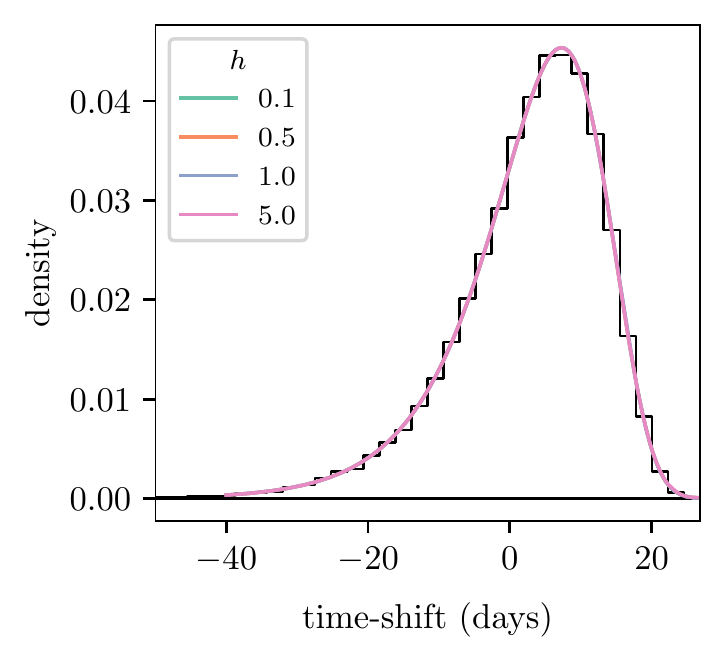}

    \caption{
        Effect of the discrete time-step, $h$, used in the embedded process. 
        The black histogram shows the empirical density and the coloured lines show the density from the PE method.  
        The computed PDFs are visually indistinguishable from one another.
    }\label{fig:effect_of_h}
\end{figure*}

We also explored the computation time of the method under the different choices of \(h\).
All tests were run on a 2021 Macbook Pro with M1 chip and the runtimes are provided in \cref{tbl:h_runtimes}.
As we increase $h$ the runtimes reduce slightly and this is a result of a smaller number of times we need to recursively evaluate \cref{eq:discrete_functional} (i.e. the value of $\kappa$ in \cref{eq:choosing_kappa}). 
The main computational expense when using the method is in the repeated inversion of LST required to evaluate the CDF and produce an approximation to it. 
This is needed for sampling realisations as would be needed within a simulation routine. 

\begin{table}[htbp]
    \centering
    \begin{tabular}{cccc}
        \toprule
        $h$ & median ($10^{-3}$~s) & mean ($10^{-3}$~s) & std ($10^{-3}$~s)\\
        \midrule
        0.1 & 13.524 & 13.537 & 0.042 \\
        0.5 & 2.974 & 3.002 & 0.061 \\
        1.0 & 1.518 & 1.517 & 0.017 \\
        5.0 & 0.419 & 0.422 & 0.012 \\
        \bottomrule
    \end{tabular}
    \caption{Benchmark results of CPU time ($10^{-3}$~s) for different choices of $h$. Results were averaged over 100 random points in the support of the distribution (i.e. $w \ge 0$).}\label{tbl:h_runtimes}
\end{table}

\subsubsection{Agreement at smaller population sizes}\label{sec:results_system_size}

Our theoretical results are valid in the limit as the population size \(N\to\infty\). 
In practical circumstances \(N\) will be finite and here we explore how the empirical distributions compare with the computed time-shift distributions in such cases. 
\Cref{fig:N_dists} shows the empirical time-shift distribution alongside the estimated distribution, using both the MM and PE methods, for different 
the population sizes (provided in the titles of each subplot).
The MM and PE methods produce PDFs that agree strongly with one another and are independent of the population size \(N\). 
Some clear deviation from the empirical time-shift distribution can be seen in the \(N = 10^3\) case but as \(N\) increases we see that the empirical distribution converges to the estimated time-shift distribution.
In the \(N = 10^4\) case there are still some minor deviations between the empirical distribution and the distributions estimated through our approaches, but the methods still appear to be approximately suitable in this situation.

\begin{figure*}[!htb]
    \centering
    \includegraphics{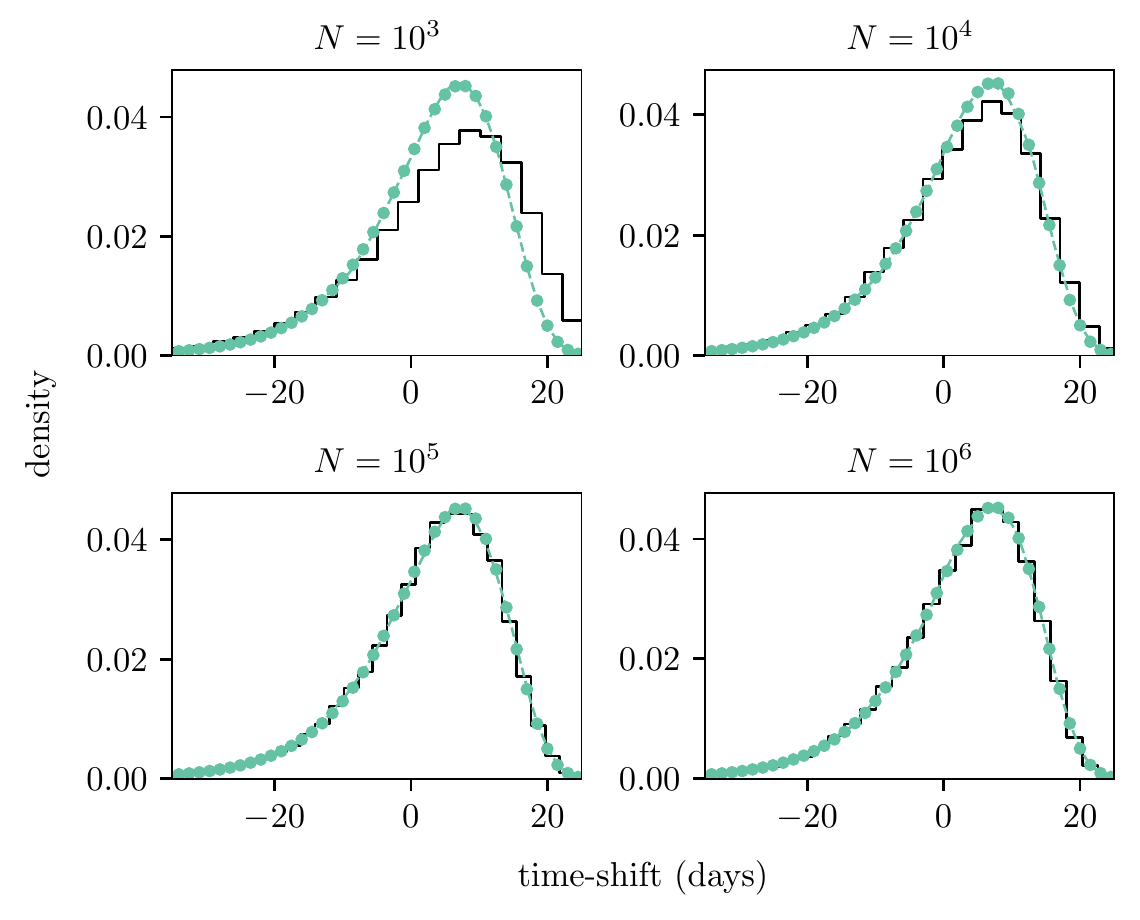}

\caption{
    PDFs for the time-shifts from the SEIR example with different population sizes (where the population size is given in the subtitle of each subplot). 
    The PDFs estimated through simulation are shown by the solid black lines. 
    PDFs for the PE and MM methods are shown by the coloured dots and {dashed lines} respectively. 
    Note that the time-shift distributions estimated using our method in each panel is independent of $N$ and thus is the same in all four panels.
}\label{fig:N_dists}
\end{figure*}

\subsubsection{Different initial conditions}\label{sec:results_initial_conditions}

In this section we briefly explore
the impact of different initial conditions on the distribution of the time-shift. 
As the LSTs for all simple initial conditions (starting with a single individual of a particular type) are computed simultaneously in \cref{alg:LST_inversion} we can easily calculate the LST for the arbitrary initial condition as detailed in \cref{sec:branching_processes}. 

\Cref{fig:z0_dists} shows the empirical time-shift distributions for varied initial conditions along the computed distributions using both the PE and MM methods.
Both MM and PE are able to reliably estimate the shape of the PDFs and capture the tail behaviours. 
An additional observation is to the shape of the time-shift distribution as the initial conditions become larger (i.e. both \(E\) and \(I\) get larger). 
When \((E, I) = (1, 0)\) the distribution is left-skewed with a median close to \(10\) and a larger variance compared to the other initial conditions. The long left tail accounts for slow growing epidemics which occur with lower probability. 
This shape aligns with intuition that stochasticity plays a larger role in an outbreak when there are few infections, which in turn influences the time taken to reach the exponential growth phase.
As we approach the \((E, I) = (15, 10)\) case, the distributions become more symmetric and the median becomes centered closer to a time-shift of \(0\) days. 
This means the model typically hits the growth phase much quicker when the initial number of individuals is larger.

\begin{figure*}[!htb]
    \centering
    \includegraphics{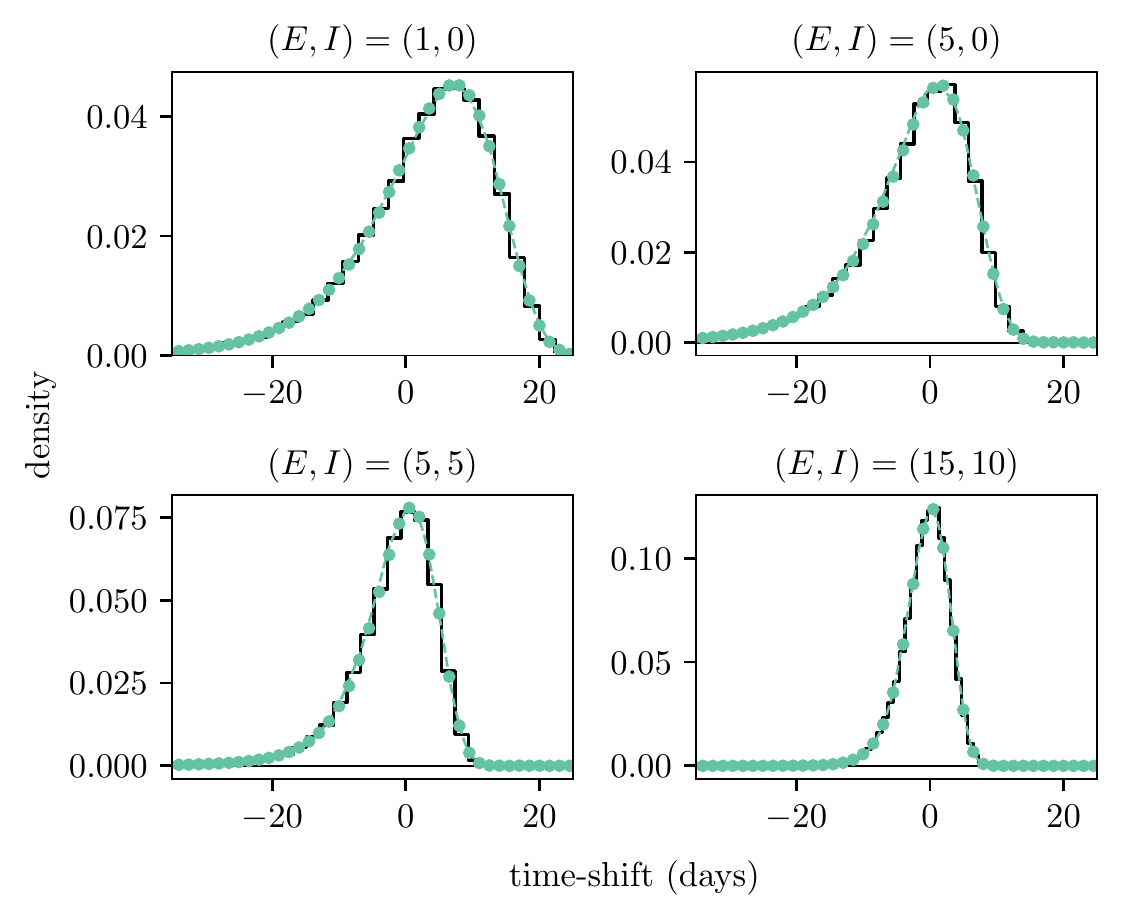}
\caption{
    PDFs for the time-shifts from the SEIR example with varying initial conditions, where the initial condition is given in the subtitle of each subplot. 
    The PDFs estimated through simulation are shown by the solid black lines. 
    PDFs for the PE and MM methods are shown by the coloured dots and {dashed lines} respectively. 
}\label{fig:z0_dists}
\end{figure*}

\subsection{Innate response model}\label{sec:innate_response_results}

Our final example is a within-host viral kinetics model that characterises a respiratory infection such as COVID-19 or influenza. 
This demonstrates our method on a model with higher complexity and shows the simplicity offered in simulating macroscopic dynamics using the time-shift approach. 
This model is a (slightly adjusted) stochastic version of that used in \citet{keVivoKineticsSARSCoV22021}, which is an extension to the so-called target-cell-limited model which incorporates an innate immune response \citep{baccamKineticsInfluenzaVirus2006}.
Our adjustment is that we also account for two mechanisms of infection, through virions and cell-to-cell interactions to further increase the complexity of the model \citep{odakaModelingViralDynamics2021}.

This model tracks \(6\) population sizes; target (or susceptible) cells (\(U\)), cells in an eclipse phase (\(E\)), infectious cells (\(I\)), number of viral particles (\(V\)), number of interferons (\(A\)) and the number of cells refractory to infection (\(R\)). 
Target (or susceptible cells) (\(U\)) can become infected when bound to by a virus particle at rate \(\beta_1\) or infected by infectious cells (\(I\)) through the viral synapse structure at rate \(\beta_2\) \citep{odakaModelingViralDynamics2021}.
We assume that contacts in these transmission processes scale with the density of cells and virus. 
We capture this by assuming some maximal carrying number of cells and virus, $K$.
Infected cells transition through an eclipse phase (\(E\)) at rate \(\sigma\) before becoming infectious (\(I\)) and being removed at rate \(\gamma\).
During the eclipse phase we account for the death of infected but not yet infectious cells at rate \(\eta\).
Over a cell's infectious period, new virus particles (\(V\)) are produced at rate \(p_v\) and are cleared at rate \(c_V\).
Furthermore, we consider the effect of an innate immune response whereby an infectious cell produces interferons (\(A\)) at rate \(p_A\) (cleared at rate \(c_A\)) which bind to cells at rate \(\delta\) and cause them to become refractory to infection \(R\).
Finally, refractory cells become target cells again at rate \(\varrho\).
This model can be formulated as a six-state CTMC with state vector \(\bs{X}(t) = (U(t), R(t), E(t), I(t), V(t), A(t))\).

Near the unstable equilibrium \(\bs{X}(0) \approx (U_0, 0, 0, 0, 0, 0)\), where \(U_0\) is the maximum number of cells lining the upper respiratory tract (URT), the process can be approximated by a CT-MBP, \(\bs{Z}(t) = (E(t), I(t), V(t))\).
This process is much simpler than the original CTMC as we don't need to directly model the immune response. 
Intuitively this is because the immune response is not critical in the early stages of an infection due to the large number of target cells.
The rates of the CTMC and CT-MBP approximation are given in \cref{tbl:flu_model_rates}.

\begin{table}[!htb]
	\centerline{\begin{tabular}{@{}cccc@{}}
			\toprule
			\(\Delta \bs{X}\)               & CTMC Rate                               & \(\Delta \bs{Z}\)         & Linearised (BP) Rate \\ \midrule
			\((U,E,V) \to (U-1, E+1, V-1)\) & \(\dfrac{\beta_1 U V}{K}\) & \(E \to E+1\)             & \(\dfrac{\beta_1 U_0}{K} V\)        \\[8pt]
			\((U,E) \to (U-1, E+1)\)        & \(\dfrac{\beta_2 U I}{K}\) & \(E \to E+1\)             & \(\dfrac{\beta_2 U_0}{K} I\)        \\[8pt]
			\((U,R) \to (U-1, R+1)\)        & \(\dfrac{\varrho U A}{K}\)  & --             & --                   \\[8pt]
			\((U,R) \to (U+1, R-1)\)        & \(\delta R\)                           & --             & --                   \\
			\((E, I) \to (E-1, I+1)\)       & \(\sigma E\)                            & \((E, I) \to (E-1, I+1)\) & \(\sigma E\)         \\
			\(E \to E-1\)                   & \(\eta E\)                              & \(E \to E-1\)             & \(\eta E\)           \\
			\(I \to I-1\)                   & \(\gamma I\)                            & \(I \to I-1\)             & \(\gamma I\)         \\
			\(V \to V+1\)                   & \(p_V I\)                               & \(V \to V+1\)             & \(p_V I\)            \\
			\(V \to V-1\)                   & \(c_V V\)                               & \(V \to V-1\)             & \(c_V V\)            \\
			\(A \to A+1\)                   & \(p_A I\)                               & --             & --                   \\
			\(A \to A-1\)                   & \(c_A A\)                               & --             & --                   \\
			\bottomrule
		\end{tabular}}
	\caption{Change in state and rates for the CTMC innate response model and the BP approximation. Note that we are overloading notation here by using the same variables for both the population numbers and the states.}
	\label{tbl:flu_model_rates}
\end{table}

Letting $\bar{\beta}_i = \beta_i U_0 / K$ for $i = 1, 2$, from \cref{tbl:flu_model_rates} the vector of lifetime parameters for the CT-MBP is 
\begin{equation*}
    \bs{a} = \left(\sigma, \gamma + p_V + \bar{\beta}_2, \bar{\beta}_1 + c_V \right)
\end{equation*}
and the PGFs are
\begin{equation*}
\begin{aligned}
	f_1(\bs{s}) & = \frac{\eta + \sigma s_2}{a_1},                               \\
	f_2(\bs{s}) & = \frac{\gamma + p_V s_2 s_3 + \bar{\beta}_2 s_2 s_1}{a_2}, \\
	f_3(\bs{s}) & = \frac{c_V + \bar{\beta}_1 s_1}{a_3}.
\end{aligned}
\end{equation*}

For the simulations we take the initial condition as \(\bs{X}_0 = (U_0-1, 0, 1, 0, 0, 0)\) where \(U_0 = 8\times 10^7\). 
We assume the maximum number of agents in the system, $K = U_0$, for convenience as this is already very large (so our approximations hold).
The parameter values used in the simulations and brief descriptions for interpretability are given in \cref{tbl:flu_model_parameters}.
We use $n = 30$ moments in the expansion and a step size of $h = 0.1$ for the PE method as this showed strong accuracy from the previous section. 
We also tested $h = 1$ but this produced divergent behaviours in the inversion (\cref{fig:TCLIR_time_shift_divergent}) which suggests that the time-scale of the process also influences the choice of $h$ and is hence model dependent.

\begin{table}[!htb]
	\centerline{\begin{tabular}{@{}ccl@{}}
			\toprule
			Parameter   & Value                  & Interpretation                                               \\ \midrule
			\(\beta_1\) & \(2.0\)                & Effective infection rate of target cells by virus particles  \\
			\(\beta_2\) & \(1.6\)                & Effective infection rate of target cells by infectious cells \\
			\(\sigma\)  & \(4.0\)                & Rate of progressing through eclipse phase                    \\
			\(\eta\)    & \(1.0\)                & Death rate of infected cells (not infectious)                \\
			\(\gamma\)  & \(1.7\)                & Recovery rate of infectious cells                            \\
			\(p_V\)     & \(45.3\)               & Production rate of virus                                     \\
			\(c_V\)     & \(10.0\)               & Clearance rate for virus                                     \\
			\(p_A\)     & \(6.0\)                & Production rate of interferons                               \\
			\(c_A\)     & \(3.0\)                & Clearance rate of interferons                                \\
			\(\varrho\) & \(104.0\)              & Effective rate of target cells becoming refractory                 \\
			\(\delta\)  & \(4.4 \times 10^{-3}\) & Rate of refractory cells returning to target           \\
			\bottomrule
		\end{tabular}}
	\caption{Parameter values and interpretations for the innate response model.}
	\label{tbl:flu_model_parameters}
\end{table}

Let \(\bs{x}(t) = K^{-1} \bs{X}(t)\) denote the density process. In the limit $K\rightarrow \infty$, a system of differential equations for the evolution of the density can be derived (see \cref{app:innate_response_derivation} for details), 
\begin{equation}\label{eq:innate_response_model}
    \begin{aligned}
        \dod{u}{t} &= -\beta_1 v u - \beta_2 i u - \varrho a u + \delta r, \\ 
        \dod{r}{t} &= \varrho a u - \delta r, \\ 
        \dod{e}{t} &= \beta_1 v u + \beta_2 i u - \sigma e - \eta e, \\ 
        \dod{i}{t} &= \sigma e - \gamma i, \\ 
        \dod{v}{t} &= p_v i - c_v v - \beta_1 u v, \\ 
        \dod{a}{t} &= p_a i - c_a a - \varrho u a.
    \end{aligned}
\end{equation}
Solving \cref{eq:innate_response_model} we obtain the deterministic approximation $\bs{X}(t) \approx K \bs{x}(t)$.

{\Cref{fig:TCLIR_example_sim} demonstrates the accurate agreement between the behaviour of a stochastic realisation and a trajectory generated using the time-shift methodology. 
The time-shift is determined from the time at which the population of virus, $V(t)$, in the stochastic realisation exceeds 2000 and thus has reached the exponential growth phase.}
The time-shift distributions themselves are shown in \cref{fig:TCLIR_time_shift} and the value of the time-shift used to produce \cref{fig:TCLIR_example_sim} (\(\tau = -0.422\)) is shown with the orange line. This example also shows that we can appropriately capture the stochastic dynamics of the full 6-state model using a simpler 3-state model \((E, I, V)\) which is then reflected in the components of the system not tracked by the BP approximation.
This in and of itself demonstrates great utility in being able to quickly generate sample paths in instances where the macroscopic dynamics are the main focus of the analysis. The agreement between the results is worse at very small population numbers, but in practice it is only once a population has grown to a large size that it even becomes observable. 

\begin{figure*}[!htb]
    \includegraphics{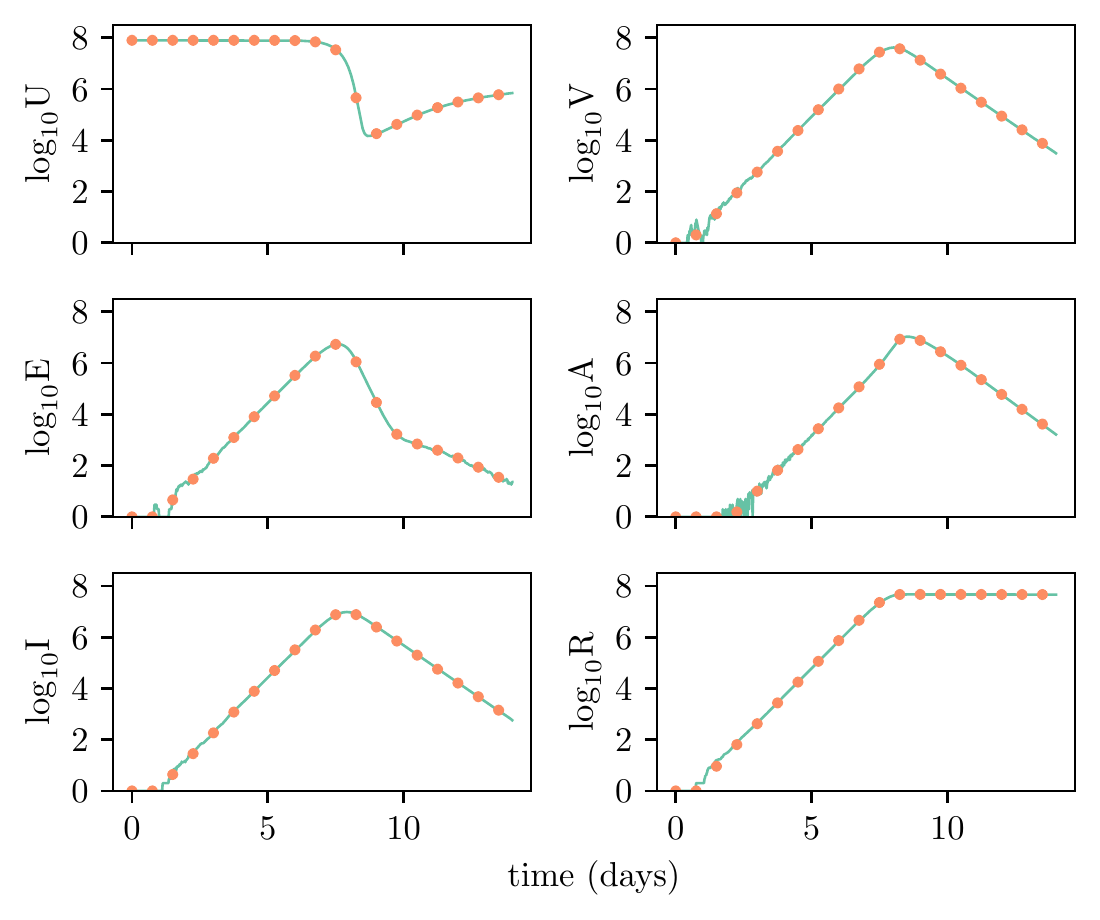}

	\caption{
        Example simulation for the innate response model. 
        The solid line indicates the stochastic simulation and the dotted line indicates the shifted deterministic solution. 
	}
	\label{fig:TCLIR_example_sim}
\end{figure*}

\begin{figure*}[!htb]
    \centering
    \includegraphics{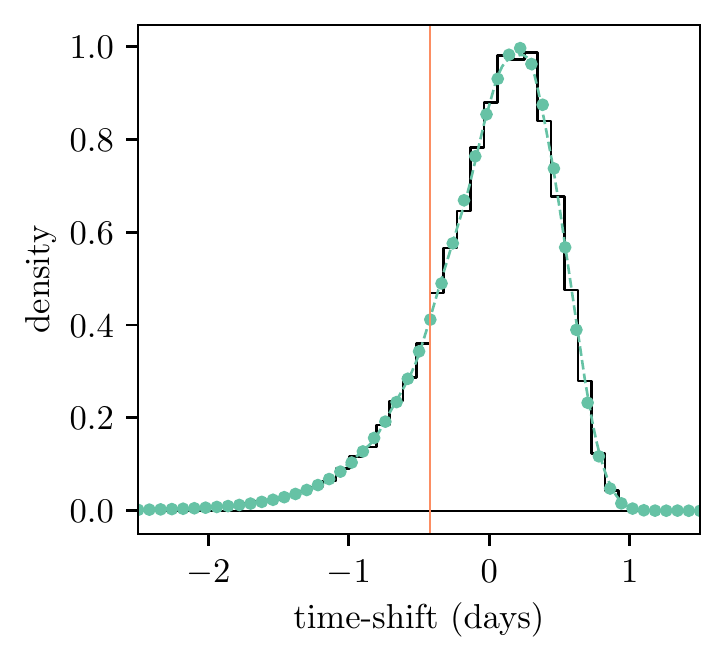}

	\caption{
        PDFs for the time-shift in the innate response model. 
        The PDF estimated through simulation is shown by the solid black lines. 
        PDFs for the PE and MM methods are shown by the coloured dots and dashed lines respectively. 
        The vertical orange line indicates the value of the time-shift used to produce \cref{fig:TCLIR_example_sim}.
	}
	\label{fig:TCLIR_time_shift}
\end{figure*}

\Cref{fig:TCLIR_time_shift} shows that both the PE and MM methods are reliably able to approximate the PDF of the time-shift. 
The time-shift distribution has lower variance compared to those seen in \cref{sec:SEIR_results}, which is a result of the faster growth and larger reproduction numbers of the populations in a within-host process. 

\section{Discussion}

We have presented a framework to {approximate} the distribution of the time-shift between sample paths of a  
broad class of Markov chain models 
and the corresponding deterministic solution. The recent work of \citet{barbourEscapeBoundaryMarkov2015} establishes a crucial theoretical result, that the initial dynamics of the Markov process affects the long-time dynamics in a manner that is largely captured by a random variable $W$ and the time-shifts is a simple transformation of this. Our contribution is a numerical framework to approximate the distribution of $W$ for the class of continuous-time Markov chains identified at the start of \cref{sec:methods}.  We introduced the PE method (\cref{sec:time_shifts}), an accurate approximation dependent on several hyper-parameters and built on the Laplace-Stieltjes transform of $W$, and the MM method (\cref{sec:moment_matching}), a fast approximation matching the first five moments of $W$ to a generalised gamma distribution.

The PE and MM methods are both flexible and can be applied to multi-type branching process models in discrete-time and continuous-time.
They allow one to generate solution curves that preserve the macroscopic behaviours of the sample paths from the stochastic model. 
Sample paths obtained in this way can be used for rapid and accurate simulation studies or inference on medium to long time scales.

The theory of branching processes is well established (see \citet{harrisBranchingProcesses1948,harrisTheoryBranchingProcesses1964,modeMultitypeBranchingProcesses1971,athreyaBranchingProcesses1972}). 
The random variable $W$ was constructed and studied in the literature as early as 1948 \citep{harrisBranchingProcesses1948,harrisMathematicalModelsBranching1951}. 
However, these bodies of work focus on models with very particular structure that aids analytic tractability; for example the linear birth-death process we derive as an early time approximation to the SIR model in \cref{sec:SIR_example} is studied in \citet{harrisMathematicalModelsBranching1951}. Another tractable example is the linear fractional branching process featured in \citet[Chapter~1]{kimmelBranchingProcessesBiology2015}.
To our knowledge, prior to this manuscript there were no tools developed specifically to compute $W$ in the more general multivariate case and our methodology provides a solution for this. 

We explored three models of varying complexity, including the SIR model in \cref{sec:SIR_example} and \cref{sec:SIR_results}, a SEIR model in \cref{sec:SEIR_results} and a 6-dimensional innate response model in \cref{sec:innate_response_results}. 
The numerical results show that both PE and MM methods reproduce the empirical distribution accurately across a broad range of model conditions, and how the parameters of the PE method can easily be chosen. 
Furthermore, the innate response model demonstrates an effective \emph{macroscale reduction} performed by our method \citep{givon_extracting_2004,roberts2015}. 
In this model, the branching process is 3-dimensional, and the computation of $W$ from this low-dimensional space allows us to accurately reproduce the macroscale dynamics of the full 6-dimensional system (shown in \cref{fig:TCLIR_example_sim}). 
This macroscale reduction is a known consequence of Theorem~1.1 of \citet{barbourEscapeBoundaryMarkov2015}, and the contribution of our manuscript is to provide a practical route to compute it.

All of the models we considered shared relatively simple branching dynamics with linear and quadratic progeny generating functions. 
This may seem restrictive, but most biological or physical processes will have dynamics of this form, especially if the common assumption of mass action mixing is made. 
The only requirement common to both the PE and MM methods is that we can compute the moments of $W$. 
It should be straightforward to extend the rules derived in \cref{sec:moments} for the calculation of the moments to more complex dynamics. 
An example exhibiting such dynamics arises in the modelling of the accumulation
of HIV-1 mutations where progeny generating functions arise that are polynomial of large (positive) integer order \citep{shiriModellingImpactAcute2011}. 
Other examples are where the offspring distribution has a parametric form, such as the negative binomial often used in epidemic models that feature super-spreading \citep{garskeEffectSuperspreadingEpidemic2008}. 
We have provided flexible open source code to automate these computations, which can be easily extended to handle new cases when they are investigated.

Our work can theoretically be extended to branching processes with non-Markovian lifetimes of individuals;
these models are referred to as age-dependent branching processes (\citealp[Chapter~IV]{athreyaBranchingProcesses1972}; \citealp[Chapter~VI]{harrisTheoryBranchingProcesses1964}; \citealp[Chapter~3]{modeMultitypeBranchingProcesses1971}).
Age-dependent processes have the same definition for $\bs{W}(t)$ (e.g. \cref{eq:Wt}) with equivalent limiting behaviours and can be studied in much the same ways as we did in \cref{sec:branching_processes}. 
The main source of difficulty in working with age-dependent processes arises from differentiating the functional equations required for computing the conditional moments of $W$. 
As opposed to the Markovian case, the integrals that need to be computed may be much more challenging to solve and may even require numerical integration to calculate. 
This would slow the computation, as exact results like those presented in \cref{sec:moments} may not arise. Current work is investigating which cases are amenable to analysis.

One of the primary advantages of our methodology is the fast generation of macroscopic sample paths that also capture the early time stochasticity.
Simulation of stochastic models in either an exact \citep{gillespieExactStochasticSimulation1977} or approximate framework \citep{gillespieApproximateAcceleratedStochastic2001} is often a computational bottleneck \citep{gillespieApproximateAcceleratedStochastic2001,blackImportanceSamplingPartially2018,kregerHybridStochasticdeterministicApproach2021}. 
This computational expense is mostly a result of the requirement to generate random numbers to calculate which event occurs next and the number of events scales linearly with the size of the system \citep{gillespieExactStochasticSimulation1977}.
Approximate stochastic simulation methods alleviate this demand somewhat by approximating the number of events over a time-step, usually referred to as tau-leaping \citep{gillespieApproximateAcceleratedStochastic2001}. 
The secondary class of methods---which our methodology can loosely be classified under---are hybrid simulation methods. 
These methods combine deterministic and stochastic simulation methods to produce more efficient simulation routines \citep{kregerHybridStochasticdeterministicApproach2021,rebuliHybridMarkovChain2017}. 
Often these methods are formulated with a regime switching during the period where the populations experience exponential growth, whereby we switch from a stochastic simulation over to a deterministic solution. 
These methods then suffer from the computational complexity required for the random number generation during the stochastic simulation regime as well as the need to solve the deterministic model with an arbitrary initial condition for each path generated \citep{rebuliHybridMarkovChain2017,kregerHybridStochasticdeterministicApproach2021}. 
Our approach generates sample paths using only a \emph{single} solve of the deterministic model and then drawing from the univariate time-shift distribution. 
Then each sample path is obtained by simply replicating the solution and shifting it by the random variate. 
This is much less computationally demanding and as such is readily applicable for inference methods that rely on forward simulations of the model such as particle filters.
{In the situation where after taking large values, the population declines back to small values, one could return to using stochastic simulations where the initial condition is determined from the deterministic solutions.}

As detailed in \cref{sec:results_hyper_params}, estimating PDF values using the PE method is fast (on the order or \(10^{-2}\)~s) and using the method we can easily draw realisations of \(W\) through standard sampling approaches (i.e. rejection sampling). 
The MM method provides further performance for sampling by approximating the distribution analytically. 
Fitting an analytical distribution means we can directly sample realisations using the inverse CDF method (taking around \(10^{-6}\)~s). 
When accuracy is paramount, the PE method offers a framework for computing the LST to arbitrary precision through increasing the number of moments and controlling the tolerance for the moment expansion in the neighbourhood of 0.
{The PE method relies on several numerical methods, each of which has associated errors.
Despite this, the main contributor to the overall error is from the error in the LST approximation in the neighbourhood of $0$ as explained in \cref{sec:time_shifts}. The errors associated with the numerical integration of the ODEs and the Laplace inversion are controlled though tolerances that can be automatically tuned in software.
}
In situations where performance is the priority then the MM method provides a much faster route to a reliable approximation of the distribution of \(W\) (as seen for the examples in \cref{sec:results}). 
However, as it is an approximation to the distribution of \(W\), it may be the case that some characteristics of the distribution are not captured (i.e. very long tails).
From our testing, matching to the first five moments yields a highly accurate approximation to the distribution. 
By providing two methods for computing the distribution of \(W\), we give a framework which can be used to assess how reliably the MM method is working for a given problem if needed.

\section{Declarations}

\textbf{Conflict of interest} \quad The authors do not have any conflicts of interest or competing interests.

\textbf{Code availability} \quad All analysis and simulations were performed using Julia 1.9 \citep{bezansonJuliaFreshApproach2017}. 
Code for producing the results and figures in this paper is available at the Github repository, \href{djmorris7/Computation_of_random_time-shifts}{https://github.com/djmorris7/Computation_of_random_time-shifts} and a package for estimating the random time-shifts is available at, \href{djmorris7/RandomTimeShifts.jl}{https://github.com/djmorris7/RandomTimeShifts.jl}.

\begin{appendices}
\appendix

\section{Time-shift distribution for the SIR example}\label{app:SIR_LST}

\Cref{eq:LST_SIR} for the LST can be inverted by applying the linearity property of inverse LSTs which allows us to invert the two terms separately. 
The inverse transform of \(q\) is simply \(q \delta(w)\) where \(\delta(w)\) is the Dirac delta function. 
This term represents the point mass at \(0\) of size \(q\) mentioned in \cref{sec:methods}. 
The inverse transform of the second term is obtained by applying the property \(\mathscr{L}^{-1}\left\{ aF(\theta) \right\}(w) = a \mathscr{L}^{-1}\left\{ F(\theta) \right\}(w)\) and noting that 
\begin{equation*}
    \mathscr{L}^{-1}\left\{ \left( 1 + \frac{\theta}{1 - q} \right)^{-1} \right\}(w) = (1-q)e^{-(1 - q)w}.
\end{equation*}
Together this yields the inverse transform
\begin{equation*}
    \mathscr{L}^{-1}\left\{ \phi(\theta) \right\}(w) = q \delta(w) + (1 - q)^2 e^{-(1 - q)w}.
\end{equation*}
We can see that the inverse transform is the sum of a point mass at \(0\) of size \(q\) and another term which corresponds to the \(w > 0\) case.
Conditioning on the event of non-extinction involves removing the point mass term at 0 and renormalising the remaining function such that it integrates to \(1\) as it is now a probability density function. 
Let \(W^\star := W \vert W > 0\) as in \cref{sec:methods}, then the density is given by 
\begin{equation}
    g_{W^\star}(w) = (1 - q) e^{-(1 - q)w}, \quad w > 0 
\end{equation}
which is precisely the density of an \(\textrm{Exp}(1-q)\) random variable and hence \(W^\star \sim \textrm{Exp}(1 - q)\).
{
Note that this result can also be arrived at if we consider the embedded discrete time process of a linear birth-death process. 
The embedded process is known to be a linear fractional process and the distribution of $W^\star$ is exponential for such processes \citep[Chapter~V]{harrisTheoryBranchingProcesses1964}.
}

The distribution of the time-shift can easily be derived by starting with the time-shift \(\tau = \lambda^{-1} (\log W - \log \mathds{E}[W])\). 
Since \(\mathds{E}[W] = 1\) when \(z_0 = 1\), we have \(\tau = \lambda^{-1} \log W\) which is equivalent to \(W = e^{\lambda \tau}\). 
Furthermore, as the transform is monotonically increasing we can obtain the density function for \(\tau^\star := \tau \vert W > 0\) (where the conditioning ensures the process doesn't fadeout) by noting that \(G_{\tau^\star}(t) = \textrm{Pr}(\tau \le t \vert W > 0) = \textrm{Pr}(W \le e^{\lambda t} \vert W > 0)\) and hence using the CDF method and differentiating we obtain
\begin{equation}
    g_{\tau^\star}(t) = \lambda (1 - q) e^{\lambda t - (1 - q) e^{\lambda t}}, \quad t \in \mathbb{R}.
\end{equation}

\section{Error calculations for LST approximation}\label{app:error_calculations_LT}

Using a Taylor series expansion on the LST
\begin{equation*}
\begin{aligned}
	\phi(\theta) & = \expect{e^{-\theta W}}                                                              \\
	             & = \mathds{E}\left[\sum_{k = 0}^{n} \frac{(-\theta W)^{k}}{k!} + R_n(\theta, s)\right] \\
	             & = \hat\phi(\theta) + \mathds{E}\left[R_n(\theta, s)\right]
\end{aligned}
\end{equation*}
where \(\hat\phi\) is from \cref{eq:laplace_transform_approximation} and \(R_n(\theta, s)\) is the Lagrange remainder term with \(s \in [0, \theta W)\). 
The remainder term is given by
\begin{equation*}
	R_n(\theta, s) = \frac{(-1)^{n+1}e^{-s}(-\theta W)^{n+1}}{(n+1)!} = \frac{e^{-s}(\theta W)^{n+1}}{(n+1)!}.
\end{equation*}
Our interest lies in bounding the error term,
\begin{equation*}
\begin{aligned}
	\mathcal{E}(s, \theta) &= \left| \mathds{E}\left[R_n(\theta, s)\right] \right| \\ 
 & \le \mathds{E}\left[\abs{ R_n(\theta, s) }\right]                                     \\
	                                                     & = \mathds{E}\left[\frac{\abs{e^{-s}} \abs{\theta^{n+1}} \abs{W^{n+1}}}{(n+1)!}\right] \\
	                                                     & = \frac{\abs{e^{-s}} \abs{\theta^{n+1}}}{(n+1)!} \mathds{E}[| W^{n+1} |]
\end{aligned}
\end{equation*}
We can simplify this by noting that \(W \ge 0\) by definition and that \(\abs{e^{-s}}\le 1\), hence
\begin{equation*}
\begin{aligned}
	\mathcal{E}(s, \theta) & \le \frac{\abs{\theta^{n+1}}}{(n+1)!} \expect{W^{n+1}} = \mathcal{E}(\theta).
\end{aligned}
\end{equation*}
Define the left-hand side of this equation as \(E_n(\theta)\). We can then determine the maximal size of the neighbourhood about 0 by solving \(E_n(\theta) < \epsilon\) for some target tolerance \(\epsilon\). For
\begin{equation*}
	\abs{\theta} \le \left(\frac{(n+1)!\epsilon}{\expect{W^{n+1}}}\right)^{1 / (n+1)}.
\end{equation*}
the error term is no more than \(\epsilon\).

\section{Step size of $h = 1$ in the innate response model}\label{app:divergent_behaviour}

\begin{figure*}[!htb]
    \centering
    \includegraphics{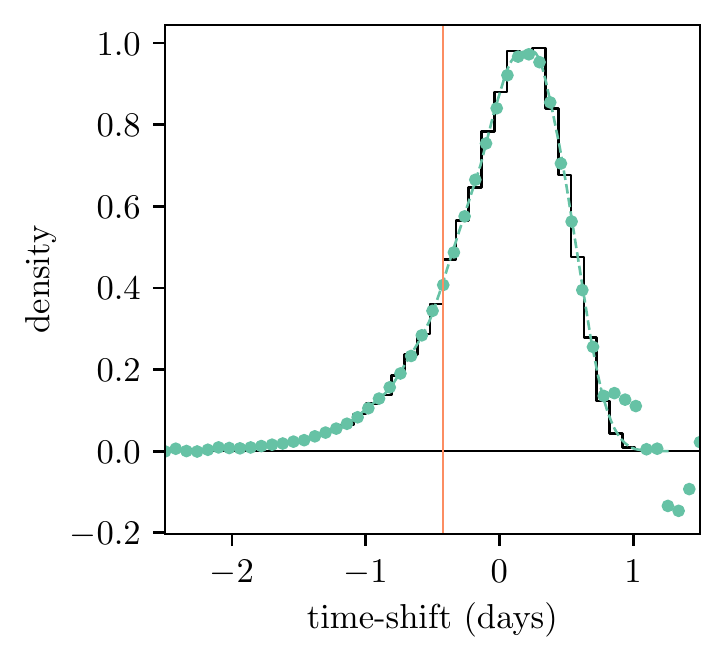}
	\caption{
        PDFs for the time-shift in the innate response model with $h = 1$ demonstrating poor numerical behaviours. 
        The PDF estimated through simulation is shown by the solid coloured lines. 
        PDFs for the PE and MM methods are shown by the coloured dots and dashed lines respectively. 
        The vertical orange line indicates the value of the time-shift used to produce \cref{fig:TCLIR_example_sim}.
	}
	\label{fig:TCLIR_time_shift_divergent}
\end{figure*}

\section{Deriving the deterministic approximation to the innate response model}\label{app:innate_response_derivation}

Let \(\bs{x}(t) = K^{-1} \bs{X}(t)\) denote the density process and note that we use lower-case letters to denote the densities (i.e. \(u = K^{-1} U\)). 
Then reading down Column 2 of \cref{tbl:flu_model_rates} and provided \(\bs{y} + \bs{l} \in \mathcal{S}\) where \(\mathcal{S}\) is the state-space, \(\boldsymbol{y}\) is the current state and \(\bs{l}\) represents the jumps of the process, then we can express the (positive) transition rates in terms of the following functions
\begin{equation}\label{eq:innate_response_functional_rates}
    \begin{aligned}
        r^{(K)}(\bs{y}, \bs{y} + \bs{l}) = \begin{cases}
            K\beta_1 u v, & \textrm{if }\bs{l} = (-1, 0, 1, 0, 1, 0) \\ 
            K\beta_2 u i, & \textrm{if }\bs{l} = (-1, 0, 1, 0, 0, 0) \\ 
            K\varrho u a, & \textrm{if }\bs{l} = (-1, 1, 0, 0, 0, 0) \\ 
            K\delta r, & \textrm{if }\bs{l} = (1, -1, 0, 0, 0, 0) \\ 
            K \sigma e, & \textrm{if }\bs{l} = (0, 0, -1, 1, 0, 0) \\ 
            K \eta e, & \textrm{if }\bs{l} = (0, 0, -1, 0, 0, 0) \\ 
            K \gamma i, & \textrm{if }\bs{l} = (0, 0, 0, -1, 0, 0) \\ 
            K p_V i, & \textrm{if }\bs{l} = (0, 0, 0, 0, 1, 0) \\ 
            K c_V v, & \textrm{if }\bs{l} = (0, 0, 0, 0, -1, 0) \\ 
            K p_A i, & \textrm{if }\bs{l} = (0, 0, 0, 0, 0, 1) \\ 
            K c_A a, & \textrm{if }\bs{l} = (0, 0, 0, 0, 0, -1).
        \end{cases}
    \end{aligned}
\end{equation}
Since the rates can all be expressed in this form (referred to as ``density dependent"), by Theorem 3.1 of \citet{kurtzSolutionsOrdinaryDifferential1970}, as \(K \to \infty\) the process \(\bs{x}(t)\) converges to \(\bs{\zeta}(t)\) (the deterministic trajectory) uniformly in probability over finite time intervals provided \(\bs{\zeta}(0) = K^{-1}\bs{X}(0)\).
{
Letting 
\begin{equation*}
    \boldsymbol{H}(\boldsymbol{\zeta}(t)) = \sum_{\boldsymbol{l}} \boldsymbol{l} r(\boldsymbol{y}, \boldsymbol{y} + \boldsymbol{l}).
\end{equation*}
where \(r(\bs{y}, \bs{y} + \bs{l}) = K^{-1} r^{(K)}(\bs{y}, \bs{y} + \bs{l})\), 
then the system of ODEs governing the deterministic approximation (\cref{eq:innate_response_model}) is given by
\begin{equation*}
    \dod{}{t}\bs{\zeta}(t) = \bs{H}(\bs{\zeta}(t)).
\end{equation*}
}
As an example, consider the first ODE \(\mathrm{d}u / \mathrm{d}t\).
The first 4 rates have non-zero first elements of \(\bs{l}\) and summing across these gives
\begin{equation*}
    \dod{u}{t} = -\beta_1 u v - \beta_2 u i - \varrho u a + \delta r
\end{equation*}
which is exactly the first differential equation in \cref{eq:innate_response_model}.
The rest of the differential equations in \cref{eq:innate_response_model} are obtained in the same way and then the deterministic trajectory is simply given by the unique solution to \cref{eq:innate_response_model} given a known initial condition \(\bar{\bs{\zeta}}(0)\).

\end{appendices}

\bibliography{references}

\begin{thebibliography}{57}
\providecommand{\natexlab}[1]{#1}
\providecommand{\url}[1]{{#1}}
\providecommand{\urlprefix}{URL }
\providecommand{\doi}[1]{\url{https://doi.org/#1}}
\providecommand{\eprint}[2][]{\url{#2}}
 \bibcommenthead

\bibitem[{Abate and Whitt(1995)}]{abateNumericalInversionLaplace1995}
Abate J, Whitt W (1995) Numerical {{Inversion}} of {{Laplace Transforms}} of {{Probability Distributions}}. ORSA Journal on Computing 7:36--43. \doi{10.1287/ijoc.7.1.36}

\bibitem[{Abate et~al.(2000)Abate, Choudhury, and Whitt}]{abateIntroductionNumericalTransform2000}
Abate J, Choudhury GL, Whitt W (2000) An {{Introduction}} to {{Numerical Transform Inversion}} and {{Its Application}} to {{Probability Models}}. In: Hillier FS, Grassmann WK (eds) Computational {{Probability}}, vol~24. {Springer US}, {Boston, MA}, p 257--323, \doi{10.1007/978-1-4757-4828-4_8}

\bibitem[{Allen(2015)}]{allenStochasticPopulationEpidemic2015}
Allen LJS (2015) Stochastic {{Population}} and {{Epidemic Models}}: {{Persistence}} and {{Extinction}}. {Springer International Publishing}, {Cham}, \doi{10.1007/978-3-319-21554-9}

\bibitem[{Allen(2017)}]{allenPrimerStochasticEpidemic2017}
Allen LJS (2017) A primer on stochastic epidemic models: {{Formulation}}, numerical simulation, and analysis. Infectious Disease Modelling 2:128--142. \doi{10.1016/j.idm.2017.03.001}

\bibitem[{Allen and Lahodny(2012)}]{allenExtinctionThresholdsDeterministic2012}
Allen LJS, Lahodny GE (2012) Extinction thresholds in deterministic and stochastic epidemic models. Journal of Biological Dynamics 6:590--611. \doi{10.1080/17513758.2012.665502}

\bibitem[{Athreya and Ney(1972)}]{athreyaBranchingProcesses1972}
Athreya KB, Ney PE (1972) Branching {{Processes}}. {Springer Berlin Heidelberg}, {Berlin, Heidelberg}, \doi{10.1007/978-3-642-65371-1}

\bibitem[{Baccam et~al.(2006)Baccam, Beauchemin, Macken, Hayden, and Perelson}]{baccamKineticsInfluenzaVirus2006}
Baccam P, Beauchemin C, Macken CA, et~al. (2006) Kinetics of {{Influenza A Virus Infection}} in {{Humans}}. Journal of Virology 80:7590--7599. \doi{10.1128/JVI.01623-05}

\bibitem[{Baker et~al.(2018)Baker, Chigansky, Hamza, and Klebaner}]{bakerPersistenceSmallNoise2018}
Baker J, Chigansky P, Hamza K, et~al. (2018) Persistence of small noise and random initial conditions. Advances in Applied Probability 50:67--81. \doi{10.1017/apr.2018.71}

\bibitem[{Barbour(1980)}]{barbourDensityDependentMarkov1980}
Barbour AD (1980) Density dependent {{Markov}} population processes. In: J{\"a}ger W, Rost H, Tautu P (eds) Biological {{Growth}} and {{Spread}}. {Springer}, {Berlin, Heidelberg}, Lecture {{Notes}} in {{Biomathematics}}, pp 36--49, \doi{10.1007/978-3-642-61850-5_4}

\bibitem[{Barbour et~al.(2015)Barbour, Hamza, Kaspi, and Klebaner}]{barbourEscapeBoundaryMarkov2015}
Barbour AD, Hamza K, Kaspi H, et~al. (2015) Escape from the boundary in {{Markov}} population processes. Advances in Applied Probability 47:1190--1211. \doi{10.1239/aap/1449859806}

\bibitem[{{Bartholomew-Biggs} et~al.(2000){Bartholomew-Biggs}, Brown, Christianson, and Dixon}]{bartholomew-biggsAutomaticDifferentiationAlgorithms2000}
{Bartholomew-Biggs} M, Brown S, Christianson B, et~al. (2000) Automatic differentiation of algorithms. Journal of Computational and Applied Mathematics 124:171--190. \doi{10.1016/S0377-0427(00)00422-2}

\bibitem[{Bauman et~al.(2023)Bauman, Chigansky, and Klebaner}]{baumanApproximationPopulationsHabitat2023}
Bauman N, Chigansky P, Klebaner F (2023) An approximation of populations on a habitat with large carrying capacity. arXiv {\href{https://arxiv.org/abs/2303.03735}{{https://arxiv.org/abs/arxiv:2303.03735}}}

\bibitem[{Baydin et~al.(2018)Baydin, Pearlmutter, Radul, and Siskind}]{baydinAutomaticDifferentiationMachine2018}
Baydin AG, Pearlmutter BA, Radul AA, et~al. (2018) Automatic {{Differentiation}} in {{Machine Learning}}: A {{Survey}}. Journal of Machine Learning Research 18:1--43

\bibitem[{Bellman and Harris(1952)}]{bellmanAgeDependentBinaryBranching1952}
Bellman R, Harris T (1952) On {{Age-Dependent Binary Branching Processes}}. The Annals of Mathematics 55:280. \doi{10.2307/1969779}

\bibitem[{Bezanson et~al.(2017)Bezanson, Edelman, Karpinski, and Shah}]{bezansonJuliaFreshApproach2017}
Bezanson J, Edelman A, Karpinski S, et~al. (2017) Julia: {{A Fresh Approach}} to {{Numerical Computing}}. SIAM Review 59:65--98. \doi{10.1137/141000671}

\bibitem[{Bishop(1996)}]{bishopNeuralNetworksPattern1996}
Bishop CM (1996) Neural {{Networks}} for {{Pattern Recognition}}, 1st edn. {Oxford University Press, USA}, {Oxford : New York}

\bibitem[{Black(2018)}]{blackImportanceSamplingPartially2018}
Black AJ (2018) Importance sampling for partially observed temporal epidemic models. Statistics and Computing 29:617--630. \doi{10.1007/s11222-018-9827-1}

\bibitem[{Black and McKane(2012)}]{blackStochasticFormulationEcological2012}
Black AJ, McKane AJ (2012) Stochastic formulation of ecological models and their applications. Trends in Ecology \& Evolution 27:337--345. \doi{10.1016/j.tree.2012.01.014}

\bibitem[{Black et~al.(2009)Black, McKane, Nunes, and Parisi}]{blackStochasticFluctuationsSusceptibleinfectiverecovered2009}
Black AJ, McKane AJ, Nunes A, et~al. (2009) Stochastic fluctuations in the susceptible-infective-recovered model with distributed infectious periods. Physical Review E, Statistical, Nonlinear, and Soft Matter Physics 80:021,922. \doi{10.1103/PhysRevE.80.021922}

\bibitem[{Black et~al.(2014)Black, House, Keeling, and Ross}]{blackEffectClumpedPopulation2014}
Black AJ, House T, Keeling MJ, et~al. (2014) The effect of clumped population structure on the variability of spreading dynamics. Journal of Theoretical Biology 359:45--53. \doi{10.1016/j.jtbi.2014.05.042}

\bibitem[{Butler and Goldenfeld(2011)}]{butlerFluctuationdrivenTuringPatterns2011}
Butler T, Goldenfeld N (2011) Fluctuation-driven {{Turing}} patterns. Physical Review E 84:011,112. \doi{10.1103/PhysRevE.84.011112}

\bibitem[{{Curran-Sebastian} et~al.(2024){Curran-Sebastian}, Pellis, Hall, and House}]{curran-sebastianCalculationEpidemicFirst2024}
{Curran-Sebastian} J, Pellis L, Hall I, et~al. (2024) Calculation of {{Epidemic First Passage}} and {{Peak Time Probability Distributions}}. SIAM/ASA Journal on Uncertainty Quantification pp 242--261. \doi{10.1137/23M1548049}

\bibitem[{Doob(1940)}]{doobRegularityPropertiesCertain1940}
Doob JL (1940) Regularity properties of certain families of chance variables. Transactions of the American Mathematical Society 47:455--486. \doi{10.1090/S0002-9947-1940-0002052-6}

\bibitem[{Dorman et~al.(2004)Dorman, Sinsheimer, and Lange}]{dormanGardenBranchingProcesses2004}
Dorman KS, Sinsheimer JS, Lange K (2004) In the {{Garden}} of {{Branching Processes}}. SIAM Review 46:202--229. {\href{https://arxiv.org/abs/20453503}{{https://arxiv.org/abs/20453503}}}

\bibitem[{Garske and Rhodes(2008)}]{garskeEffectSuperspreadingEpidemic2008}
Garske T, Rhodes CJ (2008) The effect of superspreading on epidemic outbreak size distributions. Journal of Theoretical Biology 253:228--237. \doi{10.1016/j.jtbi.2008.02.038}

\bibitem[{Gillespie(1977)}]{gillespieExactStochasticSimulation1977}
Gillespie DT (1977) Exact stochastic simulation of coupled chemical reactions. The Journal of Physical Chemistry 81:2340--2361. \doi{10.1021/j100540a008}

\bibitem[{Gillespie(2001)}]{gillespieApproximateAcceleratedStochastic2001}
Gillespie DT (2001) Approximate accelerated stochastic simulation of chemically reacting systems. The Journal of Chemical Physics 115:1716--1733. \doi{10.1063/1.1378322}

\bibitem[{Givon et~al.(2004)Givon, Kupferman, and Stuart}]{givon_extracting_2004}
Givon D, Kupferman R, Stuart A (2004) Extracting macroscopic dynamics: model problems and algorithms. Nonlinearity 17:R55--R127. \doi{10.1088/0951-7715/17/6/R01}

\bibitem[{Harris(1948)}]{harrisBranchingProcesses1948}
Harris TE (1948) Branching {{Processes}}. The Annals of Mathematical Statistics 19:474--494. \doi{10.1214/aoms/1177730146}

\bibitem[{Harris(1951)}]{harrisMathematicalModelsBranching1951}
Harris TE (1951) Some {{Mathematical Models}} for {{Branching Processes}}. Proceedings of the Second Berkeley Symposium on Mathematical Statistics and Probability 2:305--329

\bibitem[{Harris(1964)}]{harrisTheoryBranchingProcesses1964}
Harris TE (1964) The Theory of Branching Processes. {Springer-Verlag}, {Berlin}

\bibitem[{Horv{\'a}th et~al.(2020)Horv{\'a}th, Horv{\'a}th, Almousa, and Telek}]{horvathNumericalInverseLaplace2020}
Horv{\'a}th G, Horv{\'a}th I, Almousa SAD, et~al. (2020) Numerical inverse {{Laplace}} transformation using concentrated matrix exponential distributions. Performance Evaluation 137:102,067. \doi{10.1016/j.peva.2019.102067}

\bibitem[{Hubbard and Hubbard(1999)}]{hubbardVectorCalculusLinear1999}
Hubbard JH, Hubbard BB (1999) Vector Calculus, Linear Algebra, and Differential Forms: A Unified Approach. {Prentice Hall}, {Upper Saddle River, N.J}

\bibitem[{Ke et~al.(2021)Ke, Zitzmann, Ho, Ribeiro, and Perelson}]{keVivoKineticsSARSCoV22021}
Ke R, Zitzmann C, Ho DD, et~al. (2021) In vivo kinetics of {{SARS-CoV-2}} infection and its relationship with a person's infectiousness. Proceedings of the National Academy of Sciences 118:e2111477,118. \doi{10.1073/pnas.2111477118}

\bibitem[{Keeling and Rohani(2008)}]{keelingModelingInfectiousDiseases2008}
Keeling MJ, Rohani P (2008) Modeling {{Infectious Diseases}} in {{Humans}} and {{Animals}}. {Princeton University Press, New Jersey}

\bibitem[{Kendall(1966)}]{kendall:1966}
Kendall DG (1966) Branching processes since 1873. Journal of the London Mathematical Society s1-41:385--406. \doi{10.1112/jlms/s1-41.1.385}

\bibitem[{Kesten and Stigum(1966)}]{kestenLimitTheoremMultidimensional1966}
Kesten H, Stigum BP (1966) A {{Limit Theorem}} for {{Multidimensional Galton-Watson Processes}}. The Annals of Mathematical Statistics 37:1211--1223. \doi{10.1214/aoms/1177699266}

\bibitem[{Kimmel and Axelrod(2015)}]{kimmelBranchingProcessesBiology2015}
Kimmel M, Axelrod DE (2015) Branching {{Processes}} in {{Biology}}, Interdisciplinary {{Applied Mathematics}}, vol~19. {Springer New York}, {New York, NY}, \doi{10.1007/978-1-4939-1559-0}

\bibitem[{Kreger et~al.(2021)Kreger, Komarova, and Wodarz}]{kregerHybridStochasticdeterministicApproach2021}
Kreger J, Komarova NL, Wodarz D (2021) A hybrid stochastic-deterministic approach to explore multiple infection and evolution in {{HIV}}. PLOS Computational Biology 17. \doi{10.1371/journal.pcbi.1009713}

\bibitem[{Kroese et~al.(2011)Kroese, Taimre, and Botev}]{kroeseHandbookMonteCarlo2011}
Kroese DP, Taimre T, Botev ZI (2011) Handbook for {{Monte Carlo}} Methods. Wiley Series in Probability and Statistics, {Wiley}, {Hoboken, N.J}

\bibitem[{Kurtz(1970)}]{kurtzSolutionsOrdinaryDifferential1970}
Kurtz TG (1970) Solutions of ordinary differential equations as limits of pure jump markov processes. Journal of Applied Probability 7:49--58. \doi{10.2307/3212147}

\bibitem[{Kurtz(1976)}]{kurtzLimitTheoremsDiffusion1976}
Kurtz TG (1976) Limit theorems and diffusion approximations for density dependent {{Markov}} chains. In: Wets RJB (ed) Stochastic {{Systems}}: {{Modeling}}, {{Identification}} and {{Optimization}}, {{I}}. Mathematical {{Programming Studies}}, {Springer}, {Berlin, Heidelberg}, p 67--78, \doi{10.1007/BFb0120765}

\bibitem[{Ma(2020)}]{maEstimatingEpidemicExponential2020a}
Ma J (2020) Estimating epidemic exponential growth rate and basic reproduction number. Infectious Disease Modelling 5:129--141. \doi{10.1016/j.idm.2019.12.009}

\bibitem[{McKendrick(1914)}]{mckendrickStudiesTheoryContinuous1914}
McKendrick AG (1914) Studies on the {{Theory}} of {{Continuous Probabilities}}, with {{Special Reference}} to its {{Bearing}} on {{Natural Phenomena}} of a {{Progressive Nature}}. Proceedings of the London Mathematical Society s2-13:401--416. \doi{10.1112/plms/s2-13.1.401}

\bibitem[{Mode(1971)}]{modeMultitypeBranchingProcesses1971}
Mode CJ (1971) Multitype Branching Processes: Theory and Applications. Modern Analytic and Computational Methods in Science and Mathematics, {American Elsevier Pub. Co}, {New York}

\bibitem[{Nitschke et~al.(2022)Nitschke, Black, Bourrat, and Rainey}]{nitschkeEffectBottleneckSize2022}
Nitschke MC, Black AJ, Bourrat P, et~al. (2022) The {{Effect}} of {{Bottleneck Size}} on {{Evolution}} in {{Nested Darwinian Populations}}. \doi{10.1101/2022.09.22.508977}

\bibitem[{Odaka and Inoue(2021)}]{odakaModelingViralDynamics2021}
Odaka M, Inoue K (2021) Modeling viral dynamics in {{SARS-CoV-2}} infection based on differential equations and numerical analysis. Heliyon 7:e08,207. \doi{10.1016/j.heliyon.2021.e08207}

\bibitem[{Pajankar and Joshi(2022)}]{pajankarHandsonMachineLearning2022}
Pajankar A, Joshi A (2022) Hands-on {{Machine Learning}} with {{Python}}: {{Implement Neural Network Solutions}} with {{Scikit-learn}} and {{PyTorch}}. Apress, Berkeley, CA, \doi{10.1007/978-1-4842-7921-2}

\bibitem[{Rebuli et~al.(2017)Rebuli, Bean, and Ross}]{rebuliHybridMarkovChain2017}
Rebuli NP, Bean NG, Ross JV (2017) Hybrid {{Markov}} chain models of {{S}}\textendash{{I}}\textendash{{R}} disease dynamics. Journal of Mathematical Biology 75:521--541. \doi{10.1007/s00285-016-1085-2}

\bibitem[{{Revels} et~al.(2016){Revels}, {Lubin}, and {Papamarkou}}]{RevelsLubinPapamarkou2016}
{Revels} J, {Lubin} M, {Papamarkou} T (2016) Forward-mode automatic differentiation in {J}ulia. arXiv:160707892 [csMS] \urlprefix\url{https://arxiv.org/abs/1607.07892}

\bibitem[{Roberts(2015)}]{roberts2015}
Roberts A (2015) Model emergent dynamics in complex systems. SIAM, Philadelphia, PA

\bibitem[{Rogers et~al.(2012)Rogers, McKane, and Rossberg}]{rogersDemographicNoiseCan2012}
Rogers T, McKane AJ, Rossberg AG (2012) Demographic noise can lead to the spontaneous formation of species. Europhysics Letters 97:40,008. \doi{10.1209/0295-5075/97/40008}

\bibitem[{Schuster(2016)}]{schuster:2016}
Schuster P (2016) Stochasiticty in {Processes} {Fundamentals} and {Applications} to {Chemistry} and {Biology}. Springer

\bibitem[{Seneta(1981)}]{senetaNonnegativeMatricesMarkov1981}
Seneta E (1981) Non-Negative {{Matrices}} and {{Markov Chains}}. Springer {{Series}} in {{Statistics}}, {Springer}, {New York, NY}, \doi{10.1007/0-387-32792-4}

\bibitem[{Shiri and Welte(2011)}]{shiriModellingImpactAcute2011}
Shiri T, Welte A (2011) Modelling the impact of acute infection dynamics on the accumulation of {{HIV-1}} mutations. Journal of Theoretical Biology 279:44--54. \doi{10.1016/j.jtbi.2011.03.011}

\bibitem[{Turkyilmazoglu(2021)}]{turkyilmazoglu:2021}
Turkyilmazoglu M (2021) Explicit formulae for the peak time of an epidemic from the {SIR} model. Physica D: Nonlinear Phenomena 422:132,902. \doi{10.1016/j.physd.2021.132902}

\bibitem[{Wilkinson(2019)}]{wilkinsonStochasticModellingSystems2019}
Wilkinson DJ (2019) Stochastic Modelling for Systems Biology, third edition edn. Chapman \& {{Hall}}/{{CRC}} Mathematical and Computational Biology, {CRC Press, Taylor and Francis Group}, {Boca Raton}

\end{thebibliography}

\end{document}